\documentclass[a4paper,11pt]{article}

\usepackage{multirow}
\usepackage[latin1]{inputenc}
\usepackage[T1]{fontenc}
\usepackage[british]{babel}
\usepackage{graphicx}
\usepackage[dvips]{geometry}
\usepackage{float}
\usepackage{slashed}
\usepackage{pstricks,pst-fill,pst-text,pst-node,tabularx}
\usepackage{array}
\usepackage{psfrag}
\usepackage{amsmath}
\usepackage[stable]{footmisc}

\newcommand{\ket}[1]{\left|#1\right\rangle}
\newcommand{\bra}[1]{\left\langle #1\right|}
\newcommand{\Z}{\mathbf{Z}}
\newcommand{\q}{\tilde{q}}
\newcommand{\V}{\mathcal{V}}
\newcommand{\qc}{\frac{q^{-c/24}}{P(q)}}

\begin{document}

\title{Conformal two-boundary loop model on the annulus}

\author{J\'er\^ome Dubail$^{1}$, Jesper Lykke Jacobsen$^{2,1}$ and Hubert Saleur$^{1,3}$ \\[2.0mm]
  ${}^1$Institut de Physique Th\'eorique, CEA Saclay,
  91191 Gif Sur Yvette, France \\
  ${}^2$LPTENS, 24 rue Lhomond, 75231 Paris, France \\
  ${}^3$Department of Physics,
  University of Southern California, Los Angeles, CA 90089-0484}

\maketitle

\begin{abstract}

  We study the two-boundary extension of a loop model---corresponding
  to the dense phase of the O($n$) model, or to the $Q=n^2$ state
  Potts model---in the critical regime $-2 < n \le 2$. This model is
  defined on an annulus of aspect ratio $\tau$.  Loops touching the
  left, right, or both rims of the annulus are distinguished by
  arbitrary (real) weights which moreover depend on whether they wrap
  the periodic direction. Any value of these weights corresponds to a
  conformally invariant boundary condition. We obtain the exact
  seven-parameter partition function in the continuum limit, as a
  function of $\tau$, by a combination of algebraic and field
  theoretical arguments. As a specific application we derive some new
  crossing formulae for percolation clusters.

\end{abstract}

%
%
%
%

\section{Introduction}

The study of conformal boundary conditions (CBC) and boundary
operators is one of the most fruitful aspects of the vast problem of
solving two dimensional field theories and string theories. There are
many reasons for this.  In the equivalent 1+1 dimensional systems, CBC
describe possible fixed points in quantum impurity problems, such as
the multichannel Kondo problem \cite{AffleckLesHouches}, while
boundary operators decide the stability of these fixed points as well
as RG flows. In string theory, CBC describe possible branes, while RG
flows in this language decide issues of (open string) tachyon decay
\cite{SchomerusLectures}. In statistical mechanics, boundaries are
roughly where couplings to the outside take place---for instance
couplings to electrodes in quantum Hall effect type problems and their
Chalker-Coddington type lattice formulations
\cite{GruzbergLudwigRead,Cardy}.

From a more formal point of view, conformal field theories (CFTs) with
boundaries are easier to tackle than their bulk counterparts when
complicated features such as indecomposability or non-unitarity are
present. Most of the recent progress in our understanding of
logarithmic CFTs for instance has come from the consideration of their
boundary analogues \cite{PearceRasmussenZuber,ReadSaleur,Semikhatov}.

Taking a slightly different point of view, one of the basic objects in
our understanding of CFTs has been the $O(n)$ loop model, which led,
in particular, to the development of deep links with the powerful SLE
approach \cite{CardyLectures}. It is therefore no surprise that the
issue of CBC for loop models should be a major problem. This issue has
however been slow to evolve, in part for technical reasons: the
Coulomb gas formalism, which is so successful in the bulk case, is
very difficult to carry out in the presence of boundaries, for not
entirely clear reasons \cite{BauerSaleur,Cardy}. It took progress on
the algebraic side---through the study of boundary algebras and spin
models with general boundary fields---for the simplest families of CBC
to even be identified properly. The works \cite{JS1,JS2} finally
showed that CBC were obtained in the dense loop model by simply giving
to loops touching the boundary a fugacity $n_{1}$ different from the
one in the bulk. Associated conformal weights and spectra of conformal
descendents were identified, and deep connections with the blob
algebra \cite{MartinSaleur,MS2} (also called the One-Boundary
Temperley-Lieb algebra) made. Subsequently, beautiful calculations in
2D gravity \cite{Kostov,Bourgine} recovered the results of \cite{JS1,JS2}. This
will all be summarized in later sections.

Our purpose in this paper is to continue the study of \cite{JS1,JS2}
and discuss situations with several boundaries and boundary
conditions. In the case of calculations on an annulus for instance,
this means giving different weights to loops touching the left, the
right or both boundaries. We will end up in these cases with
generating functions depending on seven parameters, and of course
numerous potential applications to counting problems.

Technically, the geometrical situation on the annulus has to do with
understanding representations of Two-Boundary Temperley-Lieb algebras.
We will devote a fair amount of time to this issue, which is essential
in obtaining some of our results and conjectures. For early work and
results in this direction see \cite{deGier2BTL, Nichols}.

The problem on the annulus is also deeply related with determining the
spectra of XXZ hamiltonians with the most general boundary fields:
this has been a very active question in the Bethe ansatz community
lately \cite{Nepomechie}. We will in particular provide a complete
answer for the spectrum of these hamiltonians in the scaling limit.

More formally, the key question behind the calculations we will
present is the determination of fusion rules (and thus spectra of
boundary conditions changing operators) in loop models.  There are
deep aspects to this, some of which will be discussed here but mostly
in subsequent work.

The paper is organized as follows. At the end of this introduction, we
provide a summary of our results. Section 2 contains crucial algebraic
preliminaries, where we define and study in particular the
Two-Boundary Temperley-Lieb algebra. Section 3 contains Coulomb gas
calculations where, thanks to a realization of the boundary algebras
involving injection of charge on the boundaries, we are able to
calculate a subset of all the critical exponents of interest. This is
deeply related with the version of the problem involving XXZ chains
with boundary fields that we also discuss briefly. Section 4 is the
main section. Combining exact knowledge about hidden degeneracies
(that come in part from the algebraic analysis in Section 2---see also
\cite{DJS}), Coulomb gas arguments, and an educated guess on the
structure of boundary states, we are able to propose a formula for the
most general, seven parameters dependent partition function. Section 5
contains various combinatorial applications, and a review of the few
cases previsouly known, which our formulas all recover. In Section 6
we present a new combinatorial application, in the form of certain
refined crossing formulae for critical percolation. Finally, Section 7
gives our conclusions

\paragraph{Summary of the results:}
In this article we study a dense loop model on the annulus. Because of
the boundaries and the non-trivial topology of the annulus, there are
several types of loops, depending both on its homotopy (contractible
or not) and which boundaries (none, only left, only right, or both) it
touches.  We distinguish all these kinds of loops by giving them
different Boltzmann weights.  For convenience we always ask the number
of non-contractible lines to be even. This restriction will appear
more clearly by defining the model on a lattice in the following
section.

\begin{figure}[htbp]
\centering
a.\includegraphics[width=0.35\textwidth]{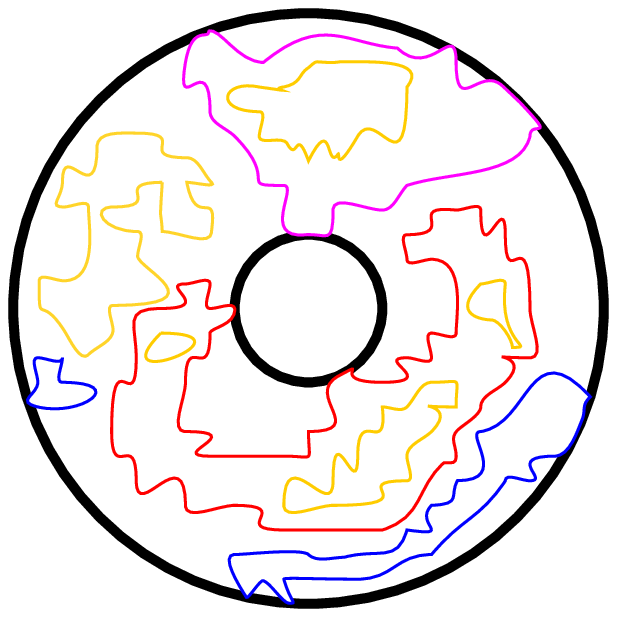} \qquad b.\includegraphics[width=0.35\textwidth]{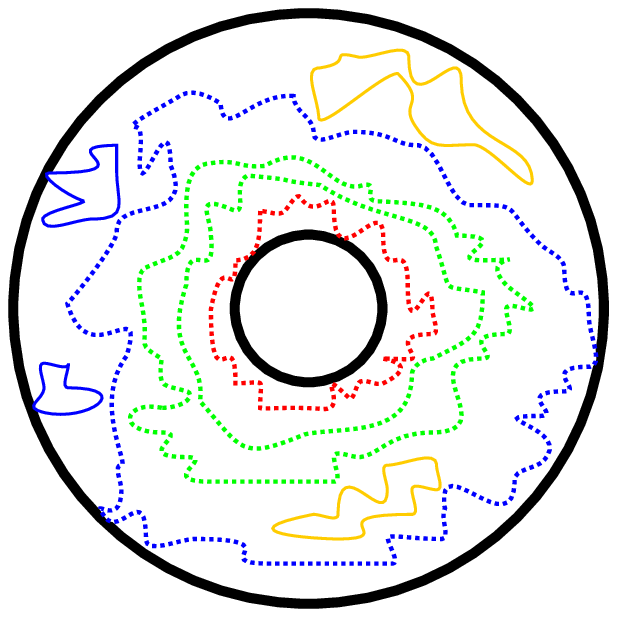}
\caption{The conformal loop model on the annulus. Different Boltzmann weights are given to the loops, depending on their topology (contractible or not) and if they touch a boundary. There can be a loop touching both boundariers if and only if there is no non-contractible loop (a). There is always an even number of non-contractible loops, and they are allowed to touch the boundaries (b).}
\label{fig:annulus}
\end{figure}

This model is endowed with conformal invariance, so we expect its
partition function to be invariant under any conformal mapping. In
particular we can study the model on a periodic strip of size $L
\times N$ ($L$ in the periodic direction), related to the annulus
$A=\left\{ z \; : \; R_1 \leq |z| \leq R_2 \right\}$ by
\begin{equation}
z' \rightarrow z=R_2 \exp \left( i 2 \pi\frac{z'}{L} \right)
\label{mapping_plane}
\end{equation}
The geometry is characterized by the modular parameter
\begin{equation}
q = e^{- \pi \tau}
\label{def:q}
\end{equation}
where $\tau = L /N = 2\pi / \log\frac{R_2}{R_1}$. As a consequence of conformal invariance, the partition function must depend on the Boltzmann weights of the loops and on the modular parameter $q$ only. It is a well-known result that the central charge of the dense loop gas is
\begin{equation}
c=1-\frac{6}{m(m+1)}
\label{param:c}
\end{equation}
where $m$ is related to the Boltzmann weight of the bulk loops $n$ through
\begin{equation}
n=2 \cos \gamma \; , \qquad \gamma = \frac{\pi}{m+1}, \qquad m>0.
\label{param:n}
\end{equation}
Note that $m$ is not restricted to be an integer. Let us also recall the Kac formula
\begin{equation}
h_{r,s} = \frac{\left[ (m+1)r -m s \right]^2-1}{4m (m+1)}.
\label{Kac}
\end{equation}

\begin{table}[htbp]
\centering
\begin{tabular}{|c|c|c|c|}
\hline
Contractible & Type & Weight & Parametrization \\
\hline
&&& \\
Yes & Bulk & $n$ & $n=2 \cos \gamma$ \\ &&& \\
Yes & Boundary $1$ & $n_1$ & $\displaystyle n_1= \frac{\sin (r_1+1) \gamma}{\sin r_1 \gamma}$, \qquad $r_1 \in ( 0,m+1 )$ \\ &&& \\
Yes & Boundary $2$ & $n_2$ & $\displaystyle n_2= \frac{\sin (r_2+1) \gamma }{\sin r_2 \gamma}$, \qquad $r_2 \in (0,m+1)$ \\ &&& \\
Yes & Both Boundaries & $n_{12}$ & $\displaystyle n_{12}= \frac{ \sin (r_1+r_2+1-r_{12}) \frac{\gamma}{2} \sin (r_1+r_2+1+r_{12}) \frac{\gamma}{2}}{\sin r_1 \gamma \sin r_2 \gamma}$ \\ &&& \\
No & Bulk & $l$ & $ \displaystyle l=2 \cos \chi$ \\ &&& \\
No & Boundary $1$ & $l_1$ & $ \displaystyle  l_1= \frac{\sin (u_1+1)\chi}{\sin u_1 \chi}$ \\ &&& \\
No & Boundary $2$ & $l_2$ & $  \displaystyle l_2= \frac{\sin (u_2+1)\chi}{\sin u_2 \chi}$ \\ &&& \\
\hline
\end{tabular}
\caption{Loop weights and their parametrizations.}
\label{param:table}
\end{table}

Now we are ready to present the main result of this article. In full generality, the partition function of the boundary loop model is given by
\begin{eqnarray}
Z &=& \frac{q^{-c/24}}{P(q)} \sum_{n \in Z} q^{h_{r_{12}-2 n,r_{12}}} \nonumber \\
&& + \frac{q^{-c/24}}{P(q)}\sum_{j \geq 1} \sum_{n \geq 0} \frac{\sin (u_1+u_2 -1 +2j) \chi  \sin \chi}{\sin u_1 \chi \sin u_2 \chi} q^{h_{r_1+r_2-1-2n, r_1+r_2-1+2j}} \nonumber \\
&& + \frac{q^{-c/24}}{P(q)}\sum_{j \geq 1} \sum_{n \geq 0} \frac{\sin (-u_1+u_2 -1 +2j) \chi  \sin \chi}{\sin -u_1 \chi \sin u_2 \chi} q^{h_{-r_1+r_2-1-2n, -r_1+r_2-1+2j}} \nonumber \\
&& + \frac{q^{-c/24}}{P(q)}\sum_{j \geq 1} \sum_{n \geq 0} \frac{\sin (u_1-u_2 -1 +2j) \chi  \sin \chi}{\sin u_1 \chi \sin -u_2 \chi} q^{h_{r_1-r_2-1-2n, r_1-r_2-1+2j}} \nonumber \\
&& + \frac{q^{-c/24}}{P(q)}\sum_{j \geq 1} \sum_{n \geq 0} \frac{\sin (-u_1-u_2 -1 +2j) \chi  \sin \chi}{\sin -u_1 \chi \sin - u_2 \chi} q^{h_{-r_1-r_2-1-2n, -r_1-r_2-1+2j}}
\label{SuperGuess}
\end{eqnarray}
where the seven parameters appearing are fixed by the seven different
loop weights. The relations between all these parameters are given in
Table~$\ref{param:table}$. Note that $P(q)$ is our notation for $\prod_{k \geq
  1} \left( 1- q^k \right)$.

%
%
%
%

\section{Some algebraic preliminaries}
\begin{figure}[htbp]
	\centering
	\includegraphics[width=0.5\textwidth]{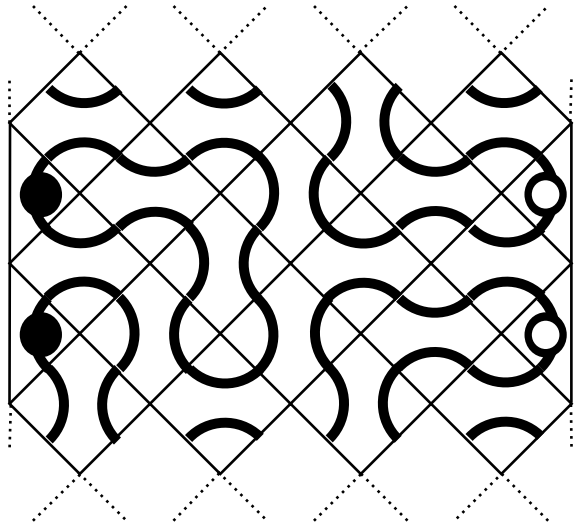}
	\caption{A configuration of dense loops on the tilted square lattice. Loops touching at least once the first (resp.\ second) boundary are marked with a black (resp.\ white) blob.}
	\label{fig:lattice_loop}
\end{figure}

Let us begin by introducing a few algebraic concepts that we will need
throughout our discussion. Our model is the densely packed loop model
on the tilted square lattice. A very convenient way to think about it
is to view it as a face model (see Fig.~$\ref{fig:lattice_loop}$). Each
face can be of two different kinds, corresponding to a horizontal or
a vertical splitting of the loops.
Each closed loop is given a Boltzmann weight $n$. The loops touching
the boundaries are distinguished from the bulk ones in our model, and
they are given different Boltzmann weights $n_1$, $n_2$ or $n_{12}$ if
they touch the first boundary, the second one, or both of them. The
total weight of a particular configuration is then $n^{N} n_1^{N_1}
n_2^{N_2} n_{12}^{N_{12}}$ where the $N_i$'s are the numbers of loops
of each kind. We shall later refine these weights to include information
about the homotopy class (contractible or not) of each loop.


\subsection{The Temperley-Lieb algebra}
To begin with, we just drop the distinction of the boundary loops.
Then partition function of such a loop model can be reformulated in
terms of local operators satisfying some commutation relations that
will correctly count the closed loops. The trick is done by the
celebrated Temperley-Lieb algebra \cite{TL}, defined as follows. The
Temperley-Lieb algebra $TL_N$ defined on $N$ strands consists of all
the words written with the $N-1$ generators $e_i$ ($1 \leq i \leq
N-1$), subject to the relations
\begin{subequations} \label{TLdef}
\begin{eqnarray}
\left|i-j \right| \geq 2 &\Rightarrow& e_i e_j=e_i e_ j\\
e_i ^2 &=& n e_i\\
e_i e_{i \pm 1} e_i &=& e_i
\end{eqnarray}
\end{subequations}
The point of this definition originates in its graphic representation.
Represent $e_i$ as an operator acting on $N$ strands
$$
\underbrace{
\begin{pspicture}(-0.1,0.)(0.7,0.8)
	\psline[](0.1,0.)(0.1,0.6)
	\psline[](0.5,0.)(0.5,0.6)
\end{pspicture}\hdots \begin{pspicture}(-0.1,0.)(0.7,0.8)
	\psellipticarc{-}(0.3,0.)(0.2,-0.25){180}{0}
	\psellipticarc{-}(0.3,0.6)(0.2,0.25){180}{0}
	\rput{0}(0.1,0.7){$^i$}
	\rput{0}(0.5,0.7){$^{i+1}$}
\end{pspicture} \hdots \begin{pspicture}(-0.1,0.)(0.7,0.8)
	\psline(0.1,0.)(0.1,0.6)
	\psline(0.5,0.)(0.5,0.6)
\end{pspicture} 
}_N
$$
then ($\ref{TLdef} b$)--($\ref{TLdef} c$) read respectively
$$
\begin{pspicture}(-0.1,0.4)(0.7,1.4)
	\psellipticarc{-}(0.3,0.)(0.2,-0.25){180}{0}
	\psellipse(0.3,0.6)(0.2,0.25)
	\psellipticarc{-}(0.3,1.2)(0.2,0.25){180}{0}
	\rput{0}(0.1,1.3){$^i$}
	\rput{0}(0.5,1.3){$^{i+1}$}
\end{pspicture} = n \begin{pspicture}(0.,0.2)(0.9,0.7)
	\psellipticarc{-}(0.3,0.)(0.2,-0.25){180}{0}
	\psellipticarc{-}(0.3,0.6)(0.2,0.25){180}{0}
	\psline(0.9,0.)(0.9,0.6)
	\rput{0}(0.1,.7){$^i$}
	\rput{0}(0.5,.7){$^{i+1}$}
\end{pspicture}
$$
\\
and
$$
\begin{pspicture}(-0.1,0.7)(1.,1.9)
	\psellipticarc{-}(0.3,0.)(0.2,-0.25){180}{0}
	\psellipticarc{-}(0.3,0.6)(0.2,0.25){180}{0}
	\psellipticarc{-}(0.7,0.6)(0.2,-0.25){180}{0}
	\psellipticarc{-}(0.7,1.2)(0.2,0.25){180}{0}
	\psellipticarc{-}(0.3,1.2)(0.2,-0.25){180}{0}
	\psellipticarc{-}(0.3,1.8)(0.2,0.25){180}{0}
	\psline(0.1,0.6)(0.1,1.2)
	\psline(0.9,0.)(0.9,0.6)
	\psline(0.9,1.2)(0.9,1.8)
	\rput{0}(0.1,1.9){$^i$}
	\rput{0}(0.5,1.9){$^{i+1}$}
\end{pspicture} = \begin{pspicture}(0.,0.2)(1.0,0.7)
	\psellipticarc{-}(0.3,0.)(0.2,-0.25){180}{0}
	\psellipticarc{-}(0.3,0.6)(0.2,0.25){180}{0}
	\psline(0.9,0.)(0.9,0.6)
	\rput{0}(0.1,.7){$^i$}
	\rput{0}(0.5,.7){$^{i+1}$}
\end{pspicture}.
$$
\\
\\
Each configuration of loops on a lattice of width $N$ can be written
as a particular word of the algebra $TL_N$ (for example the
configuration in Fig.~$\ref{fig:lattice_loop}$, dropping the blobs
coming from the boundaries, would be written $e_1 e_3 e_7 e_6 e_2 e_1
e_5 e_7 e_6 e_3 e_7$). In fact all the configurations can be generated
by taking powers of the transfer matrix of the model, which reads
\begin{equation}
T'_N= \left( \prod_{\substack{i=1 \\ i \; \mathrm{odd}}}^{N} 1+e_i \right) \left( \prod_{\substack{i=1 \\ i \; \mathrm{even}}}^{N} 1+e_i\right).
\label{0BTL_transfer}
\end{equation}


\subsection{Boundary conditions and blob operators}
In the model we have just introduced, the loops touching the left or
right boundaries of the lattice are not different from the other ones.
We will refer to this particularly simple case as ``free'' boundary
conditions. In this paper we deal with much more general boundary
conditions. They consist in giving a different Boltzmann weight $n_1$
(resp.\ $n_2$) to the loops which have touched at least once the
boundary $1$ (resp.\ $2$). This is encoded in the transfer matrix by
the addition of so-called ``blob'' operators $b_1$ and $b_2$ to the
algebra $TL_N$. Their graphical representation consists of a black
(resp.\ white) blob which marks the first (resp.\ last) strand. $b_1$
acts as
$$
\underbrace{
\begin{pspicture}(-0.1,0.)(0.7,0.8)
	\psline(0.1,0.)(0.1,0.6)
	\psline(0.5,0.)(0.5,0.6)
	\pscircle*(0.1,0.3){0.1}
\end{pspicture}\hdots \begin{pspicture}(-0.1,0.)(0.7,0.8)
	\psline(0.1,0.)(0.1,0.6)
	\psline(0.5,0.)(0.5,0.6)
\end{pspicture} 
}_N
$$
and $b_2$ as
$$
\underbrace{
\begin{pspicture}(-0.1,0.)(0.7,0.8)
	\psline(0.1,0.)(0.1,0.6)
	\psline(0.5,0.)(0.5,0.6)
\end{pspicture}\hdots \begin{pspicture}(-0.1,0.)(0.7,0.8)
	\psline(0.1,0.)(0.1,0.6)
	\psline(0.5,0.)(0.5,0.6)
	\pscircle(0.5,0.3){0.1}
\end{pspicture} 
}_N.
$$
They satisfy the defining relations
\begin{subequations} \label{blobdef1}
\begin{eqnarray}
i\geq 2 &\Rightarrow& b_1 e_i = e_i b_1\\
b_1^2 &=& b_1\\
e_1 b_1 e_1&=& n_1 e_1 
\end{eqnarray}
\end{subequations}
and
\begin{subequations} \label{blobdef2}
\begin{eqnarray}
i\leq N-2 &\Rightarrow& b_2 e_i = e_i b_2\\
b_2^2&=&b_2 \label{blobdef5}\\
e_{N-1} b_2 e_{N-1}&=& n_2 e_{N-1}
\end{eqnarray}
\end{subequations}
In what follows, we will assume that $N$ is always even. In that case it is possible to have closed loops touching both boundaries. In order to count each of these loops with a weight $n_{12}$, we impose the relation
\begin{equation}
\left( \prod_{\substack{i=1 \\ i \; \mathrm{even}}}^{N} e_i \right) b_1 b_2 \left( \prod_{\substack{i=1 \\ i \; \mathrm{odd}}}^{N} e_i\right)  \left( \prod_{\substack{i=1 \\ i \; \mathrm{even}}}^{N} e_i\right) =  n_{12} \left( \prod_{\substack{i=1 \\ i \; \mathrm{even}}}^{N} e_i\right)
\label{2BTL_quotient}
\end{equation}
which can be drawn as
$$
\begin{pspicture}(-0.1,0.7)(1.8,1.8)
	\psline(0.1,0.6)(0.1,1.2)
	\pscircle*(0.1,0.9){0.1}
	\psellipticarc{-}(0.3,0.)(0.2,-0.25){180}{0}
	\psellipticarc{-}(0.3,0.6)(0.2,0.25){180}{0}
	\psellipticarc{-}(1.1,0.)(0.2,-0.25){180}{0}
	\psellipticarc{-}(1.1,0.6)(0.2,0.25){180}{0}
	\psellipticarc{-}(0.7,0.6)(0.2,-0.25){180}{0}
	\psellipticarc{-}(0.7,1.2)(0.2,0.25){180}{0}
	\psellipticarc{-}(1.5,0.6)(0.2,-0.25){180}{0}
	\psellipticarc{-}(1.5,1.2)(0.2,0.25){180}{0}
	\psellipticarc{-}(0.3,1.2)(0.2,-0.25){180}{0}
	\psellipticarc{-}(0.3,1.8)(0.2,0.25){180}{0}
	\psellipticarc{-}(1.1,1.2)(0.2,-0.25){180}{0}
	\psellipticarc{-}(1.1,1.8)(0.2,0.25){180}{0}
\end{pspicture}\hdots \begin{pspicture}(-0.1,0.7)(1.8,1.8)
	\psline(1.5,0.6)(1.5,1.2)
	\pscircle(1.5,0.9){0.1}
	\psellipticarc{-}(1.3,0.)(0.2,-0.25){180}{0}
	\psellipticarc{-}(1.3,0.6)(0.2,0.25){180}{0}
	\psellipticarc{-}(0.5,0.)(0.2,-0.25){180}{0}
	\psellipticarc{-}(0.5,0.6)(0.2,0.25){180}{0}
	\psellipticarc{-}(0.9,0.6)(0.2,-0.25){180}{0}
	\psellipticarc{-}(0.9,1.2)(0.2,0.25){180}{0}
	\psellipticarc{-}(0.1,0.6)(0.2,-0.25){180}{0}
	\psellipticarc{-}(0.1,1.2)(0.2,0.25){180}{0}
	\psellipticarc{-}(1.3,1.2)(0.2,-0.25){180}{0}
	\psellipticarc{-}(1.3,1.8)(0.2,0.25){180}{0}
	\psellipticarc{-}(0.5,1.2)(0.2,-0.25){180}{0}
	\psellipticarc{-}(0.5,1.8)(0.2,0.25){180}{0}
\end{pspicture} = n_{12} \begin{pspicture}(-0.1,0.2)(1.4,0.6)
	\psellipticarc{-}(0.3,0.)(0.2,-0.25){180}{0}
	\psellipticarc{-}(0.3,0.6)(0.2,0.25){180}{0}
	\psellipticarc{-}(1.1,0.)(0.2,-0.25){180}{0}
	\psellipticarc{-}(1.1,0.6)(0.2,0.25){180}{0}
\end{pspicture} \hdots \begin{pspicture}(0.,0.2)(1.3,0.6)
	\psellipticarc{-}(0.3,0.)(0.2,-0.25){180}{0}
	\psellipticarc{-}(0.3,0.6)(0.2,0.25){180}{0}
	\psellipticarc{-}(1.1,0.)(0.2,-0.25){180}{0}
	\psellipticarc{-}(1.1,0.6)(0.2,0.25){180}{0}
\end{pspicture}.
$$
\\
\\
The generators $e_i$, $b_1$ and $b_2$, subject to the relations
(\ref{TLdef}), (\ref{blobdef1})--(\ref{blobdef2}) and the quotient
(\ref{2BTL_quotient}), thus form the Two-Boundary Temperley-Lieb
algebra on $N$ strands, denoted $2BTL_N$. A simpler case to which we
shall sometimes refer is the One-Boundary Temperley-Lieb algebra
$1BTL_N$, generated only by the $e_i$'s and $b_1$. The transfer matrix
of the two-boundary loop model is then a generalization of
Eq.~($\ref{0BTL_transfer}$)
\begin{equation}
T_N= b_1 b_2 \left( \prod_{\substack{i=1 \\ i \; \mathrm{odd}}}^{N} 1+e_i \right) \left( \prod_{\substack{i=1 \\ i \; \mathrm{even}}}^{N} 1+e_i\right).
\label{2BTL_transfer}
\end{equation}
It generates all the boundary loop configurations on a strip (see
Fig.~$\ref{fig:lattice_loop}$) and gives the correct weights $n$ to the
closed loops in the bulk, and $n_1$, $n_2$ or $n_{12}$ to the ones
touching the boundaries.


\subsection{Generic irreducible representations of 2BTL}
Irreducible representations of the Temperley-Lieb algebra are well
known, and are closely related to those of the quantum group
$SU(2)_q$. When $q$ is not a root of unity\footnote{Of course here $q$
  is not the modular parameter defined by ($\ref{def:q}$).}, the
representation theory of $SU(2)_q$ is essentially the same as the one
of $SU(2)$. In that case, the corresponding irreducible
representations of the Temperley-Lieb algebra are said to be generic.
The generic representations have a simple graphical interpretation, as
the Temperley-Lieb algebra itself. The different modules
(representation spaces) $\V_s$ are given by configurations of
$(N-s)/2$ half-loops and $s$ strings. For example, consider the
Temperley-Lieb algebra on $4$ strands $TL_4$, for which there are only
three generic modules.
$$
\V_0 = \left\{ \begin{array}{c}
\begin{pspicture}(0.,0.)(1.3,0.7)
	\psellipticarc{-}(.3,0.5)(0.2,0.5){180}{0}
	\psellipticarc{-}(1.1,0.5)(0.2,0.5){180}{0}
\end{pspicture}\\ 
\begin{pspicture}(0.,0.)(1.3,0.7)
	\psellipticarc{-}(.7,0.5)(0.2,0.3){180}{0}
	\psellipticarc{-}(.7,0.5)(0.6,0.5){180}{0}
\end{pspicture} 
\end{array}  \right\}
\qquad \V_2 = \left\{ \begin{array}{c}
\begin{pspicture}(0.,0.)(1.3,0.7)
	\psellipticarc{-}(1.1,0.5)(0.2,0.5){180}{0}
	\psline(0.1,0.)(0.1,0.5)
	\psline(0.5,0.)(0.5,0.5)
\end{pspicture}\\ 
\begin{pspicture}(0.,0.)(1.3,0.7)
	\psellipticarc{-}(.7,0.5)(0.2,0.5){180}{0}
	\psline(0.1,0.)(0.1,0.5)
	\psline(1.3,0.)(1.3,0.5)
\end{pspicture}\\ 
\begin{pspicture}(0.,0.)(1.3,0.7)
	\psellipticarc{-}(.3,0.5)(0.2,0.5){180}{0}
	\psline(0.9,0.)(0.9,0.5)
	\psline(1.3,0.)(1.3,0.5)
\end{pspicture} 
\end{array}  \right\} \qquad \V_4 = \left\{ \begin{array}{c}
\begin{pspicture}(0.,0.)(1.3,0.6)
	\psline(0.1,0.)(0.1,0.5)
	\psline(0.5,0.)(0.5,0.5)
	\psline(0.9,0.)(0.9,0.5)
	\psline(1.3,0.)(1.3,0.5)
\end{pspicture}
\end{array}  \right\}
$$
The vertical lines are the strings, and the action of $e_i$ on two
strings on the sites $i$, $i+1$ is defined to be zero.

Now if we work with the Two-Boundary Temperley-Lieb algebra $2BTL_N$
(or with $1BTL_N$), the generic representation theory is quite similar
and has been studied in \cite{MartinSaleur,JS2,deGier2BTL}. The
modules consist of the all states formed with half-loops and strings,
but the half-loops can be marked with black or white blobs. Note that
every black blob is necessarily on the left of every white blob. One
can show \cite{JS2,deGier2BTL,DJS} that the dimension of $\V_0$ for
$2BTL_N$ is $2^N$. For example, the module $\V_0$ is of dimension $16$
for $2BTL_4$:
$$
\V_0 = \left\{ \begin{array}{ccccccc}
\begin{pspicture}(0.,0.)(1.3,0.7)
	\psellipticarc{-}(.3,0.5)(0.2,0.5){180}{0}
	\psellipticarc{-}(1.1,0.5)(0.2,0.5){180}{0}
\end{pspicture} &, &\begin{pspicture}(0.,0.)(1.3,0.7)
	\pscircle*(0.15,0.25){0.1}
	\psellipticarc{-}(.3,0.5)(0.2,0.5){180}{0}
	\psellipticarc{-}(1.1,0.5)(0.2,0.5){180}{0}
\end{pspicture} &, & \begin{pspicture}(0.,0.)(1.3,0.7)
	\pscircle(1.25,0.25){0.1}
	\psellipticarc{-}(.3,0.5)(0.2,0.5){180}{0}
	\psellipticarc{-}(1.1,0.5)(0.2,0.5){180}{0}
\end{pspicture} &, & \begin{pspicture}(0.,0.)(1.3,0.7)
	\pscircle*(0.15,0.25){0.1}
	\pscircle(1.25,0.25){0.1}
	\psellipticarc{-}(.3,0.5)(0.2,0.5){180}{0}
	\psellipticarc{-}(1.1,0.5)(0.2,0.5){180}{0}
\end{pspicture}\\
\begin{pspicture}(0.,0.)(1.3,0.7)
	\pscircle*(0.95,0.25){0.1}
	\psellipticarc{-}(.3,0.5)(0.2,0.5){180}{0}
	\psellipticarc{-}(1.1,0.5)(0.2,0.5){180}{0}
\end{pspicture} &, &\begin{pspicture}(0.,0.)(1.3,0.7)
	\pscircle(.45,0.25){0.1}
	\psellipticarc{-}(.3,0.5)(0.2,0.5){180}{0}
	\psellipticarc{-}(1.1,0.5)(0.2,0.5){180}{0}
\end{pspicture} &, &\begin{pspicture}(0.,0.)(1.3,0.7)
	\pscircle*(0.15,0.25){0.1}
	\pscircle(0.45,0.25){0.1}
	\psellipticarc{-}(.3,0.5)(0.2,0.5){180}{0}
	\psellipticarc{-}(1.1,0.5)(0.2,0.5){180}{0}
\end{pspicture} &, &\begin{pspicture}(0.,0.)(1.3,0.7)
	\pscircle*(0.95,0.25){0.1}
	\pscircle(1.25,0.25){0.1}
	\psellipticarc{-}(.3,0.5)(0.2,0.5){180}{0}
	\psellipticarc{-}(1.1,0.5)(0.2,0.5){180}{0}
\end{pspicture} \\
\begin{pspicture}(0.,0.)(1.3,0.7)
	\pscircle*(.15,0.25){0.1}
	\pscircle*(0.95,0.25){0.1}
	\psellipticarc{-}(.3,0.5)(0.2,0.5){180}{0}
	\psellipticarc{-}(1.1,0.5)(0.2,0.5){180}{0}
\end{pspicture} &, & \begin{pspicture}(0.,0.)(1.3,0.7)
	\pscircle(0.45,0.25){0.1}
	\pscircle(1.25,0.25){0.1}
	\psellipticarc{-}(.3,0.5)(0.2,0.5){180}{0}
	\psellipticarc{-}(1.1,0.5)(0.2,0.5){180}{0}
\end{pspicture} &, & \begin{pspicture}(0.,0.)(1.3,0.7)
	\pscircle*(.15,0.25){0.1}
	\pscircle(.45,0.25){0.1}
	\pscircle(1.25,0.25){0.1}
	\psellipticarc{-}(.3,0.5)(0.2,0.5){180}{0}
	\psellipticarc{-}(1.1,0.5)(0.2,0.5){180}{0}
\end{pspicture} &, & \begin{pspicture}(0.,0.)(1.3,0.7)
	\pscircle*(0.15,0.25){0.1}
	\pscircle*(.95,0.25){0.1}
	\pscircle(1.25,0.25){0.1}
	\psellipticarc{-}(.3,0.5)(0.2,0.5){180}{0}
	\psellipticarc{-}(1.1,0.5)(0.2,0.5){180}{0}
\end{pspicture}\\
\begin{pspicture}(0.,0.)(1.3,0.7)
	\psellipticarc{-}(.7,0.5)(0.2,0.3){180}{0}
	\psellipticarc{-}(.7,0.5)(0.6,0.5){180}{0}
\end{pspicture} &, & \begin{pspicture}(0.,0.)(1.3,0.7)
	\psellipticarc{-}(.7,0.5)(0.2,0.3){180}{0}
	\psellipticarc{-}(.7,0.5)(0.6,0.5){180}{0}
	\pscircle*(0.2,0.25){0.1}
\end{pspicture} &, & \begin{pspicture}(0.,0.)(1.3,0.7)
	\psellipticarc{-}(.7,0.5)(0.2,0.3){180}{0}
	\psellipticarc{-}(.7,0.5)(0.6,0.5){180}{0}
	\pscircle(1.2,0.25){0.1}
\end{pspicture} &, & \begin{pspicture}(0.,0.)(1.3,0.7)
	\psellipticarc{-}(.7,0.5)(0.2,0.3){180}{0}
	\psellipticarc{-}(.7,0.5)(0.6,0.5){180}{0}
	\pscircle*(0.2,0.25){0.1}
	\pscircle(1.2,0.25){0.1}
\end{pspicture} 
\end{array}  \right\}
$$
This result will play an important role in the sequel, when we will
have to deal with Coulomb gas arguments.

Now consider the modules with strings. Half-loops between strings
cannot be blobbed, since they are always separated from the boundary
by at least one string, so $b_1$ or $b_2$ cannot act on them. The
leftmost (resp.\ rightmost) string can carry a black (resp.\ white)
blob. They can also be orthogonal to the blob, in the sense that they
are not eigenstates of the projector $b_1$, but of the orthogonal
projector (``unblob'') $1-b_1$. Let us thus mark with a black (resp.\
white) square the action of $1-b_1$ (resp.\ $1-b_2$). Then there is
not only one module with $s$ strings, but four, depending on the blob
status (blobbed or unblobbed) of the leftmost and rightmost strings.
For $2BTL_4$ the modules with two strings are
$$
\V^{bb}_2 = \left\{ \begin{array}{c}
\begin{pspicture}(0.,0.)(1.3,0.7)
	\psline(0.1,0.)(0.1,0.5)
	\psline(0.5,0.)(0.5,0.5)
	\pscircle*(0.1,0.25){0.1}
	\pscircle(0.5,0.25){0.1}
	\psellipticarc{-}(1.1,0.5)(0.2,0.5){180}{0}
\end{pspicture} \\
\begin{pspicture}(0.,0.)(1.3,0.7)
	\psline(0.1,0.)(0.1,0.5)
	\psline(0.5,0.)(0.5,0.5)
	\pscircle*(0.1,0.25){0.1}
	\pscircle(0.5,0.25){0.1}
	\pscircle(1.25,0.25){0.1}
	\psellipticarc{-}(1.1,0.5)(0.2,0.5){180}{0}
\end{pspicture} \\
\begin{pspicture}(0.,0.)(1.3,0.7)
	\psline(0.1,0.)(0.1,0.5)
	\psline(1.3,0.)(1.3,0.5)
	\pscircle*(0.1,0.25){0.1}
	\pscircle(1.3,0.25){0.1}
	\psellipticarc{-}(.7,0.5)(0.2,0.5){180}{0}
\end{pspicture} \\
\begin{pspicture}(0.,0.)(1.3,0.7)
	\psline(0.9,0.)(0.9,0.5)
	\psline(1.3,0.)(1.3,0.5)
	\pscircle*(0.9,0.25){0.1}
	\pscircle(1.3,0.25){0.1}
	\psellipticarc{-}(0.3,0.5)(0.2,0.5){180}{0}
\end{pspicture}\\
\begin{pspicture}(0.,0.)(1.3,0.7)
	\psline(0.9,0.)(0.9,0.5)
	\psline(1.3,0.)(1.3,0.5)
	\pscircle*(0.9,0.25){0.1}
	\pscircle(1.3,0.25){0.1}
	\pscircle*(0.15,0.25){0.1}
	\psellipticarc{-}(0.3,0.5)(0.2,0.5){180}{0}
\end{pspicture}
\end{array}  \right\} \quad \V^{bu}_2 = \left\{ \begin{array}{c}
\begin{pspicture}(0.,0.)(1.3,0.7)
	\psline(0.1,0.)(0.1,0.5)
	\psline(0.5,0.)(0.5,0.5)
	\pscircle*(0.1,0.25){0.1}
	\psframe(0.4,0.15)(0.6,0.35)
	\psellipticarc{-}(1.1,0.5)(0.2,0.5){180}{0}
\end{pspicture} \\
\begin{pspicture}(0.,0.)(1.3,0.7)
	\psline(0.1,0.)(0.1,0.5)
	\psline(0.5,0.)(0.5,0.5)
	\pscircle*(0.1,0.25){0.1}
	\psframe(0.4,0.15)(0.6,0.35)
	\pscircle(1.25,0.25){0.1}
	\psellipticarc{-}(1.1,0.5)(0.2,0.5){180}{0}
\end{pspicture} \\
\begin{pspicture}(0.,0.)(1.3,0.7)
	\psline(0.1,0.)(0.1,0.5)
	\psline(1.3,0.)(1.3,0.5)
	\pscircle*(0.1,0.25){0.1}
	\psframe(1.2,0.15)(1.4,0.35)
	\psellipticarc{-}(.7,0.5)(0.2,0.5){180}{0}
\end{pspicture} \\
\begin{pspicture}(0.,0.)(1.3,0.7)
	\psline(0.9,0.)(0.9,0.5)
	\psline(1.3,0.)(1.3,0.5)
	\pscircle*(0.9,0.25){0.1}
	\psframe(1.2,0.15)(1.4,0.35)
	\psellipticarc{-}(0.3,0.5)(0.2,0.5){180}{0}
\end{pspicture}\\
\begin{pspicture}(0.,0.)(1.3,0.7)
	\psline(0.9,0.)(0.9,0.5)
	\psline(1.3,0.)(1.3,0.5)
	\pscircle*(0.15,0.25){0.1}
	\pscircle*(0.9,0.25){0.1}
	\psframe(1.2,0.15)(1.4,0.35)
	\psellipticarc{-}(0.3,0.5)(0.2,0.5){180}{0}
\end{pspicture}
\end{array}  \right\} \quad \V^{ub}_2 = \left\{ \begin{array}{c}
\begin{pspicture}(0.,0.)(1.3,0.7)
	\psline(0.1,0.)(0.1,0.5)
	\psline(0.5,0.)(0.5,0.5)
	\psframe*(0.,0.15)(0.2,0.35)
	\pscircle(0.5,0.25){0.1}
	\psellipticarc{-}(1.1,0.5)(0.2,0.5){180}{0}
\end{pspicture} \\
\begin{pspicture}(0.,0.)(1.3,0.7)
	\psline(0.1,0.)(0.1,0.5)
	\psline(0.5,0.)(0.5,0.5)
	\psframe*(0.,0.15)(0.2,0.35)
	\pscircle(0.5,0.25){0.1}
	\pscircle(1.25,0.25){0.1}
	\psellipticarc{-}(1.1,0.5)(0.2,0.5){180}{0}
\end{pspicture} \\
\begin{pspicture}(0.,0.)(1.3,0.7)
	\psline(0.1,0.)(0.1,0.5)
	\psline(1.3,0.)(1.3,0.5)
	\psframe*(0.,0.15)(0.2,0.35)
	\pscircle(1.3,0.25){0.1}
	\psellipticarc{-}(.7,0.5)(0.2,0.5){180}{0}
\end{pspicture} \\
\begin{pspicture}(0.,0.)(1.3,0.7)
	\psline(0.9,0.)(0.9,0.5)
	\psline(1.3,0.)(1.3,0.5)
	\psframe*(0.8,0.15)(1.,0.35)
	\pscircle(1.3,0.25){0.1}
	\psellipticarc{-}(0.3,0.5)(0.2,0.5){180}{0}
\end{pspicture}\\
\begin{pspicture}(0.,0.)(1.3,0.7)
	\psline(0.9,0.)(0.9,0.5)
	\psline(1.3,0.)(1.3,0.5)
	\psframe*(0.8,0.15)(1.,0.35)
	\pscircle*(0.15,0.25){0.1}
	\pscircle(1.3,0.25){0.1}
	\psellipticarc{-}(0.3,0.5)(0.2,0.5){180}{0}
\end{pspicture}
\end{array}  \right\} \quad \V^{uu}_2 = \left\{ \begin{array}{c}
\begin{pspicture}(0.,0.)(1.3,0.7)
	\psline(0.1,0.)(0.1,0.5)
	\psline(0.5,0.)(0.5,0.5)
	\psframe*(0.,0.15)(0.2,0.35)
	\psframe(0.4,0.15)(0.6,0.35)
	\psellipticarc{-}(1.1,0.5)(0.2,0.5){180}{0}
\end{pspicture} \\
\begin{pspicture}(0.,0.)(1.3,0.7)
	\psline(0.1,0.)(0.1,0.5)
	\psline(0.5,0.)(0.5,0.5)
	\psframe*(0.,0.15)(0.2,0.35)
	\psframe(0.4,0.15)(0.6,0.35)
	\pscircle(1.25,0.25){0.1}
	\psellipticarc{-}(1.1,0.5)(0.2,0.5){180}{0}
\end{pspicture} \\
\begin{pspicture}(0.,0.)(1.3,0.7)
	\psline(0.1,0.)(0.1,0.5)
	\psline(1.3,0.)(1.3,0.5)
	\psframe*(0.,0.15)(0.2,0.35)
	\psframe(1.2,0.15)(1.4,0.35)
	\psellipticarc{-}(.7,0.5)(0.2,0.5){180}{0}
\end{pspicture} \\
\begin{pspicture}(0.,0.)(1.3,0.7)
	\psline(0.9,0.)(0.9,0.5)
	\psline(1.3,0.)(1.3,0.5)
	\psframe*(0.8,0.15)(1.,0.35)
	\psframe(1.2,0.15)(1.4,0.35)
	\psellipticarc{-}(0.3,0.5)(0.2,0.5){180}{0}
\end{pspicture}\\
\begin{pspicture}(0.,0.)(1.3,0.7)
	\psline(0.9,0.)(0.9,0.5)
	\psline(1.3,0.)(1.3,0.5)
	\psframe*(0.8,0.15)(1.,0.35)
	\pscircle*(0.15,0.25){0.1}
	\psframe(1.2,0.15)(1.4,0.35)
	\psellipticarc{-}(0.3,0.5)(0.2,0.5){180}{0}
\end{pspicture}
\end{array}  \right\}.
$$
Obviously these modules are related to each other by the blobbed/unblobbed transformations
\begin{equation}
b_{1,2} \rightarrow 1 - b_{1,2}
\label{blobbed_unblobbed}
\end{equation}
and this symmetry between the projectors $b_1$ and $b_2$ and their orthogonals $1-b_1$ and $1-b_2$ will indeed play some role in our analysis of the loop model.


\subsection{Markov trace on 2BTL and the boundary loop model on the annulus}

Let us begin by dropping the blobs and the particular boundary
weights, and start with the free/free partition function. Then the
transfer matrix is given by ($\ref{0BTL_transfer}$) in terms of the
Temperley-Lieb generators $e_i$'s. There is an algebraic tool closely
linked with the study of the Temperley-Lieb algebra, namely the Markov
trace, which is useful for our problem. We give only a naive
definition of that tool here. The Markov trace of an element $M$ of
the Temperley-Lieb algebra is the number which is obtained by
identifying the top and the bottom of the diagramatic representation
of $M$, upon counting each closed loop with a weight $n$. For example,
consider $M=e_1 e_2$ on $N=4$ strands:
$$
\mathrm{Tr} \left\{ e_1 e_2 \right\} = \mathrm{Tr} \left\{ \begin{pspicture}(0.,0.2)(1.4,0.7)
	\psline(0.9,0.)(0.1,0.6)
	\psline(1.3,0.)(1.3,0.6)
	\psellipticarc{-}(0.7,0.6)(0.2,0.25){180}{0}
	\psellipticarc{-}(0.3,0.)(0.2,0.25){0}{180}
\end{pspicture} \right\}= n^2.
$$
Although this object clearly depends on the number of strands $N$, we
do not mention it explicitly. Given this naive definition, the Markov
trace is associated with the Temperley-Lieb algebra itself, and does
not require to know anything about its representations. Note that the
Markov trace is not a trace in the common sense: it is not a sum over
a basis of states of the diagonal action of $M$. However, there is a
well-known relation between the Markov trace and the usual traces over
the different generic modules of the Temperley-Lieb algebra. Using
$n=2 \cos \gamma$, we have
\begin{equation}
\mathrm{Tr} M = \sum_{\substack{s \geq 0}} \frac{\sin (s+1) \gamma}{\sin \gamma} \; tr_{\V_s} M
\label{MarkovOBTL}.
\end{equation}
This relation calls for a few remarks. First, we use the convention
that $\V_s$ is empty if $s>N$ or if $s \neq N \; mod \; 2$. In
particular, the sum is finite, and the terms depend on $N$. Then, note
that $\frac{\sin (s+1) \gamma}{\sin \gamma}$ is a polynomial in the
variable $n$ : $\frac{\sin 2 \gamma}{\sin \gamma}=n$, $\frac{\sin 3
  \gamma}{\sin \gamma}=n^2-1$, $\frac{\sin 4 \gamma}{\sin
  \gamma}=n^3-2n$, etc. They are actually Chebyshev polynomials of the
second kind, $U_s(n/2)$.

\smallskip

We consider now a modification of the Markov trace, which will be
useful for our loop model. We can decide that we draw all the
Temperley-Lieb diagrams in $\mathbf{R} - \left\{ 0\right\}$ instead of
$\mathbf{R}$ and that when we compute the Markov trace, we give a
weight $n$ only to the contractible loops, and another weight $l=2
\cos \chi$ to the non-contractible ones. Again it is convenient to
consider an example:
$$
\mathrm{Tr}_{\chi} \left\{\begin{pspicture}(-0.1,0.2)(1.4,0.7)
	\psline[linewidth=1pt](0.1,0.0)(.1,0.6)
	\psline[linewidth=1pt](0.5,0.0)(0.5,0.6)
	\psellipticarc[showpoints=false,arrowscale=2]{-}(1.1,0.6)(0.2,0.25){180}{0}
	\psellipticarc[showpoints=false,arrowscale=2]{-}(1.1,0.)(0.2,0.25){0}{180}
\end{pspicture} \right\} = \begin{pspicture}(-0.1,0.2)(3.6,2.)
	\psline[linewidth=1pt](0.1,0.0)(.1,0.6)
	\psline[linewidth=1pt](0.5,0.0)(0.5,0.6)
	\psellipticarc[showpoints=false,arrowscale=2]{-}(1.1,0.6)(0.2,0.25){180}{0}
	\psellipticarc[showpoints=false,arrowscale=2]{-}(1.1,0.)(0.2,0.25){0}{180}
	\pscircle*(1.72,0.3){0.1}
	\rput[b]{0}(1.88,0.47){$0$}
	\psellipticarc[linestyle=dashed]{-}(1.72,0.3)(0.51,0.8){-151}{151}	
	\psellipticarc[linestyle=dashed]{-}(1.72,0.3)(0.89,1.1){-158}{158}	
	\psellipticarc[linestyle=dashed]{-}(1.72,0.3)(1.32,1.35){-160}{160}	
	\psellipticarc[linestyle=dashed]{-}(1.72,0.3)(1.7,1.7){-164}{164}	
\end{pspicture} = l^2 n .
$$
\\
\\
\\
\\
The modified Markov trace has quite the same structure as the previous
one. In particular, it can be decomposed on the usual traces over the
generic modules exactly in the same way.
\begin{equation}
\mathrm{Tr}_\chi M = \sum_{\substack{s \geq 0}} \frac{\sin (s+1) \chi}{\sin \chi} \; tr_{\V_s} M
\label{ModifiedMarkovOBTL}.
\end{equation}
This time the coefficient of each trace is a polynomial in $l$, and it
is a remarkable fact that it does not depend on $n$ at all.

\smallskip

It turns out that we can define the Markov trace (or the modified
Markov trace) in the same way for the Two-Boundary Temperley-Lieb
algebra $2BTL$, counting the blobbed loops with the appropriate weight
$n_1$, $n_2$ or $n_{12}$ (or $l_1$, $l_2$ for non-contractible loops
in the case of the modified Markov trace). Recall that contractible
loops touching both boundaries appear only if we work with an even
number of strands $N$, so the number of strings must always be even,
and we write $s=2j$. Again, this object admits a decomposition on the
usual traces over the different generic modules
\begin{equation}
\mathrm{Tr}_\chi M = tr_{\V_0}M + \sum_{\substack{j \geq 1 \\ \alpha,\beta = b,u}} D^{\alpha \beta}_{2j} \; tr_{\V^{\alpha \beta}_{2j}} M
\label{ModifiedMarkov2BTL}.
\end{equation}
where the $D_{2j}^{\alpha \beta}$ are some polynomials in $n_1$, $n_2$
and $n$ (or $l_1$, $l_2$, $l$ only if we are dealing with the modified
Markov trace). The computation of the coefficients $D^{\alpha
  \beta}_{2j}$ can be achieved by various methods, see \cite{JS2} for
a combinatorial proof or \cite{DJS} for a more algebraic approach. The
results are as follows. Let $u_1$ and $u_2$ be such that
\begin{equation}
l_1 = \frac{\sin (u_1+1) \chi}{\sin u_1 \chi} 
\label{param:l1}
\end{equation}
and
\begin{equation}
l_2 = \frac{\sin (u_2+1) \chi}{\sin u_2 \chi} 
\label{param:l2}
\end{equation}
then
\begin{subequations}
\begin{eqnarray}
D^{bb}_{2j} &=& \frac{\sin (u_1+u_2-1+2j) \chi \sin \chi}{\sin u_1 \chi \sin u_2 \chi}\\
D^{bu}_{2j} &=& \frac{\sin (u_1-u_2-1+2j) \chi \sin \chi}{\sin u_1 \chi \sin -u_2 \chi}\\
D^{ub}_{2j} &=& \frac{\sin (-u_1+u_2-1+2j) \chi \sin \chi}{\sin -u_1 \chi \sin u_2 \chi}\\
D^{uu}_{2j} &=& \frac{\sin (-u_1-u_2-1+2j) \chi \sin \chi}{\sin -u_1 \chi \sin -u_2 \chi}.
\end{eqnarray}
\label{amplitudes}
\end{subequations}
These equations are related by the blobbed/unblobbed transformation ($\ref{blobbed_unblobbed}$). To see this, note that the weight of a non-contractible loop marked with a black square (recall the black square stands for the action of $1-b_1$) is simply $l-l_1 = \frac{\sin(- u_1 +1)\gamma}{\sin -u_1 \gamma}$. The transformation ($\ref{blobbed_unblobbed}$) has thus the effect of changing $u_1$ into $-u_1$, or $u_2$ into $-u_2$. These are indeed the transformations needed to pass from $D_{2j}^{bb}$ to $D_{2j}^{ub}$, or $D_{2j}^{bu}$, or $D^{uu}_{2j}$.

\smallskip

The physical interest of the Markov trace, or of the modified Markov
trace, is that it counts automatically with the correct weight all the
loops of a Temperley-Lieb element when the top and the bottom of the
diagram are identified. This is exactly what we need to write down the
partition function of our loop model. The transfer matrix on $N$ strands
is an element of the algebra $2BTL_N$, see ($\ref{2BTL_transfer}$). We
want to work on an annulus of size $L \times N$, so taking periodic
boundary conditions in the $L$ direction, the partition function of
our loop model is just the modified Markov trace of a power of the
transfer matrix.
\begin{equation}
Z=\mathrm{Tr}_\chi T_{N}^L
\label{Markov_transfer}
\end{equation}
Eq.~($\ref{ModifiedMarkov2BTL}$) gives the natural decomposition over
the different modules
\begin{equation}
Z = tr_{\V_0} T_N^L + \sum_{\substack{j \geq 1 \\ \alpha,\beta = b,u}} D^{\alpha \beta}_{2j} \; tr_{\V^{\alpha \beta}_{2j}} T_N^L
\label{Z_MM2BTL}.
\end{equation}
This relation holds for every $N$ and $L$. In particular, it must
remain true in the limit $L,N \rightarrow \infty$ with $L/N$ fixed.
Then if we introduce the (properly renormalized) characters
\begin{equation}
K^{\alpha \beta}_{2j} = \left\{ \lim_{L,N \rightarrow \infty} tr_{\V^{\alpha \beta}_{2j}} T_N^L \right\}_{\mathrm{renorm.}}
\end{equation}
the conformal partition function will have the following structure
\begin{equation}
Z=K_0 + \sum_{\substack{j \geq 1 \\ \alpha, \beta =b,u}} D^{\alpha\beta}_{2j} K^{\alpha \beta}_{2j}.
\label{FormSuperGuess}
\end{equation}
Hence, the computation of the conformal partition function has been
reduced to the determination of the characters $K^{\alpha
  \beta}_{2j}$.

%
%
%
%

\section{Coulomb gas for the sector without strings}

In the previous section we explained why the partition function should
have an algebraic structure coming from the Two-Boundary
Temperley-Lieb algebra that we have just presented, and hence can be
decomposed on different sectors corresponding to the generic
irreducible representations of $2BTL$. Hence we are allowed to deal
with each sector independently. This section is devoted to the
computation by Coulomb gas arguments of the character $K_0$, defined
in the previous section as the trace over the module $\V_0$ of
$2BTL_N$. First we detail how to obtain the parametrizations given in
Table~$\ref{param:table}$ in the Coulomb gas framework. These
parametrizations have also a deeper algebraic origin associated with
the Temperley-Lieb algebra \cite{MartinSaleur,JS2,deGier2BTL,DJS}, but
the detailed discussion of this aspect will be deferred to \cite{DJS}.


\subsection{A reminder: Coulomb gas on an infinite cylinder}

We now want to map our boundary loop model on a height model for which
it is simpler to compute some quantities such as correlation functions
or partition functions. We recall some classical arguments here. For
the loop model on an infinite cylinder, the mapping is well-known.
This must correspond to the limit $N \gg L$ (recall $L$ is the
periodic direction). First begin by giving each loop an orientation,
then interpret the oriented loops as level lines for a height field
$h$ defined on the cylinder. The height varies by $\Delta h = \pm \pi$
when upon crossing an oriented line. Each loop is counted with a
weight $e^{\pm i \gamma}$, depending on its orientation. The sum over
the two orientations then gives the correct initial weight $n =
e^{i\gamma} + e^{-i\gamma}$ to the original loop. Then it is generally
argued that this model renormalizes to a free gaussian model with
action
\begin{equation}
\mathcal{S} = \frac{g}{4 \pi} \int \left( \partial h \right)^2 \, {\rm d}^2 x.
\label{action:coulomb}
\end{equation}
This is however not sufficient to count correctly the loops which
wrap around the cylinder. To do this, one has to add two charges
$e^{\pm i (\gamma/\pi ) h}$ at the ends of the cylinder. This modifies
the scaling dimension of the vertex operator $e^{i \alpha h}$ to
\begin{equation}
\Delta_\alpha = \frac{g}{4}\left\{\left( \alpha +\gamma/\pi \right)^2 - \left( \gamma/\pi \right)^2\right\}.
\end{equation}
The value of $g$ can then be fixed by the following argument. We
started from a model in which the height difference when passing
through a loop is $\Delta h = \pm \pi$, so the operator $\cos 2 h$
should be marginal. This recquires $\Delta_2 = 2$ or $\Delta_{-2}=2$,
so
\begin{equation}
g=1 \pm \frac{\gamma}{\pi}.
\end{equation}
The choice of the sign can actually lead to two different phases of
the loop model, dense or dilute. We are working with a dense loop
model, so we have to choose the solution $g<1$. To finish, let us
determine the central charge of this conformal field theory. The
addition of charges at the ends of the cylinder has changed the
behaviour of the partition function on the very long cylinder ($N \gg
L$) by a factor $e^{\pi N (\gamma/\pi)^2 / g}$. This is sufficient to
identify the central charge, since we expect $Z \sim e^{-\pi c N /
  6L}$ in that limit, instead of $e^{- \pi N/6L}$ without the addition
of charges. Then we have
\begin{equation}
c = 1-6\frac{(\gamma/\pi)^2}{g}.
\end{equation}
Defining $m$ such that $\gamma = \frac{\pi}{m+1}$, this is nothing but the well-known formula ($\ref{param:c}$).

\subsection{Boundaries in the height model}
\begin{figure}[htbp]
	\centering
	\psfrag{eq}[c]{$=$}
	\psfrag{plus}[c]{$+$}
	\psfrag{alpha}[l]{$+e^{i\gamma}$}
	\psfrag{beta}[l]{$+e^{-i\gamma}$}
	\includegraphics[width=0.7\textwidth]{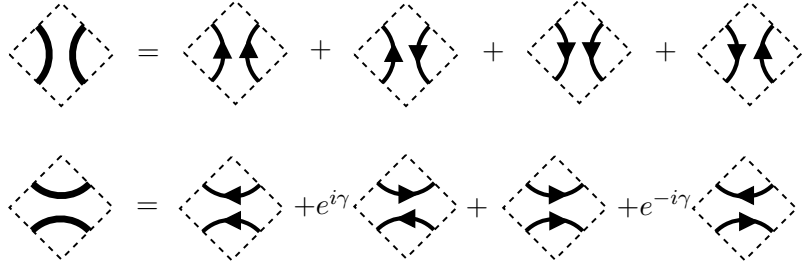}
	\caption{Faces for the oriented loop model used for the Coulomb gas construction. The first line is just the identity in the Temperley-Lieb algebra, while the second line is a generator $e_i$ satisfying ($\ref{TLdef}$).}
	\label{fig:TLorient}
\end{figure}

Now we turn to the finite geometry of the annulus, and deal with the
boundaries. Begin again by giving an orientation to each loop. The
bulk loops are counted with a weight $e^{-i\gamma}$ if they are
clockwise oriented, and $e^{i \gamma}$ in the other case. The
Temperley-Lieb generators $e_i$'s hence can be defined as acting on
the orientations as shown in Fig.~$\ref{fig:TLorient}$. It is not
difficult to check that the $e_i's$ satisfy the relations
($\ref{TLdef}$) also in the oriented loop language.

\begin{figure}[htbp]
	\centering
	\psfrag{alpha1}[l]{$\displaystyle = \frac{1}{2 i \sin r_1 \gamma} \left\{ \begin{array}{l} \\ \\ \\  \\ \end{array}\right.$}
	\psfrag{alpha2}[l]{$- e^{-ir_1 \gamma}$}
	\psfrag{alpha3}[l]{$+i e^{-i r_{12}\gamma}$}
	\psfrag{alpha4}[l]{$+ e^{ir_1 \gamma}$}
	\psfrag{alpha5}[l]{$+i e^{i r_{12} \gamma}$}
	\psfrag{alpha6}[l]{$\displaystyle \left. \begin{array}{l} \\ \\ \\   \\ \end{array} \right\} $}
	\psfrag{beta1}[l]{$\displaystyle = \frac{1}{2 i \sin r_2 \gamma} \left\{ \begin{array}{l} \\ \\ \\  \\ \end{array}\right.$}
	\psfrag{beta2}[l]{$ e^{ir_2 \gamma}$}
	\psfrag{beta3}[l]{$+ \; i $}
	\psfrag{beta4}[l]{$-e^{-i r_2 \gamma}$}
	\psfrag{beta5}[l]{$+ \; i$}
	\psfrag{beta6}[l]{$\displaystyle \left. \begin{array}{l} \\ \\ \\   \\ \end{array} \right\} $}
	\includegraphics[width=0.85\textwidth]{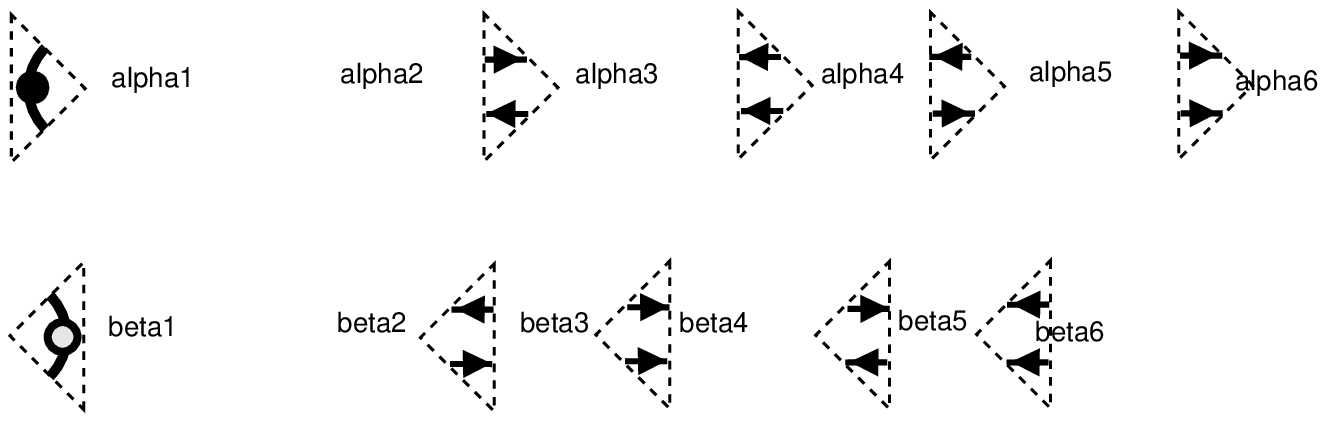}
	\caption{Action of the blobs on the oriented loops. The orientation of the loops is not conserved by the blobs.}
	\label{fig:TLbloborient}
\end{figure}

We must find out how the blob operator $b_1$ acts on the loop
orientation. There are four different faces (triangles) with half
oriented loops which can be combined to create $b_1$ (see
Fig.~$\ref{fig:TLbloborient}$). Two of them conserve the orientation
of the loop, which means that there is one arrow coming from the left
side of the triangle, and one arrow entering it. The two others do not
conserve it: both arrows point in the same direction. It is clear that
the two faces which do not conserve the orientation cannot contribute
to the weight of a loop touching only this boundary, because the
orientation will be conserved everywhere else, so when we close the
loop this contribution just vanishes. Now assume that the blob just
adds some arbitrary phase factor $e^{\pm i r_1 \gamma}$ to a closed
loop. Then requiring ($\ref{blobdef1} c$), the loop gets a weight $n_1
\propto \sin (r_r +1)\gamma$ instead of $n= 2 \cos \gamma$. The
correct normalization is fixed by ($\ref{blobdef1} b$). There remains
one free parameter: the phase of the coefficients of the faces which
do not conserve the orientation. We end up with the expression of the
blob $b_1$ given in Fig.~$\ref{fig:TLbloborient}$, where $e^{i r_{12}
  \gamma}$ is our free parameter.

\smallskip

\begin{figure}[htbp]
	\centering
	\psfrag{eq}[l]{$\displaystyle = \; \frac{-1}{4 \sin r_1 \gamma \sin r_2 \gamma}$}
	\psfrag{alpha2}[l]{$e^{-i(r_1+r_2+1) \gamma}$}
	\psfrag{alpha3}[l]{$-  e^{-i r_{12}\gamma}$}
	\psfrag{alpha4}[l]{$+  e^{i(r_1+r_2+1) \gamma}$}
	\psfrag{alpha5}[l]{$- \; e^{ir_{12} \gamma}$}
	\psfrag{par1}[l]{$\left\{ \begin{array}{c} \\ \\ \\ \\ \\ \\ \end{array} \right.$}
	\psfrag{par2}[l]{$\left. \begin{array}{c} \\ \\ \\ \\ \\ \\ \end{array} \right\}$}
	\includegraphics[width=0.85\textwidth]{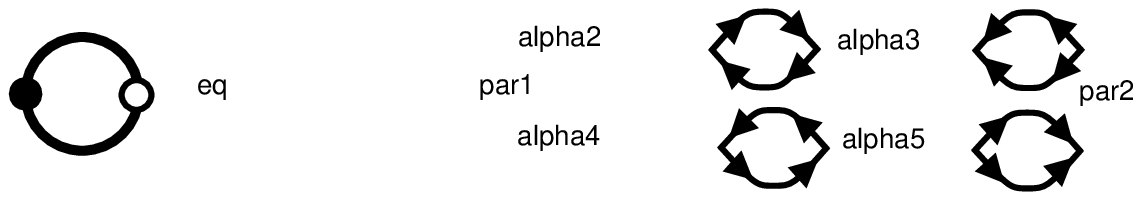}
	\caption{The four terms giving the parametrization ($\ref{param:nb}$) for the weight of a loop touching both boundaries.}
	\label{fig:coulomb_weight}
\end{figure}

The same can be done for the second blob $b_2$, so we actually have
two free parameters coming from the boundary faces which do not
conserve the orientation. Our problem has a global phase invariance,
so one of them can be fixed, to give the expression of $b_2$ shown in
Fig.~$\ref{fig:TLbloborient}$. All the different loop weights can then
be computed in terms of the parameters $r_1$, $r_2$ and $r_{12}$. The
weights $n_1$ and $n_2$ are given by the sum over both orientations
\begin{equation}
n_1 = \frac{\sin (r_1 +1) \gamma}{\sin r_1 \gamma}
\end{equation}
and
\begin{equation}
n_2 = \frac{\sin (r_2 +1) \gamma}{\sin r_2 \gamma}.
\end{equation}
The weight of a loop touching both boundaries is a sum over four
possible configurations of the orientations (see
Fig.~$\ref{fig:coulomb_weight}$), giving the parametrization
\begin{equation}
n_{12}=\frac{\sin \left(\displaystyle  \frac{r_1+r_2+1+r_{12}}{2} \gamma \right) \sin \left( \displaystyle \frac{r_1+r_2+1-r_{12}}{2}\gamma \right)}{\sin r_1 \gamma \sin r_2 \gamma}
\label{param:nb}
\end{equation}
as claimed in the introduction (see Table~$\ref{param:table}$).

\subsection{Spectrum in the sector without strings}

In the sector without strings, we can compute the conformal character
$K_0$ using Coulomb gas arguments. The previous prescription for the
operators $b_1$ and $b_2$ gives us a height model with the action
($\ref{action:coulomb}$), with Neumann boundary conditions $\partial_y
h (x,y=0) = \partial_y h (x,y=N)=0$. Because of the boundary vertices
which introduce some magnetic charge in the system, we see that there
can be a difference of height if we turn once around the annulus:
$$
h(x+L,y)= 2p \pi  + h(x,y) , \qquad p \in \Z.
$$
Clearly, $p$ is the number of boundary vertices which inject charge in
the system minus the number of those which take charge from it (see
Fig.~$\ref{fig:TLbloborient}$). Such a configuration must be counted
with a weight $e^{i p r_{12} \gamma}$.

\begin{figure}[htbp]
\centering
\psfrag{alpha1}[l]{$x=0$}
\psfrag{alpha2}[l]{$x=L$}
\psfrag{beta1}[l]{$y=0$}
\psfrag{beta2}[l]{$y=N$}
\qquad \includegraphics[width=0.7\textwidth]{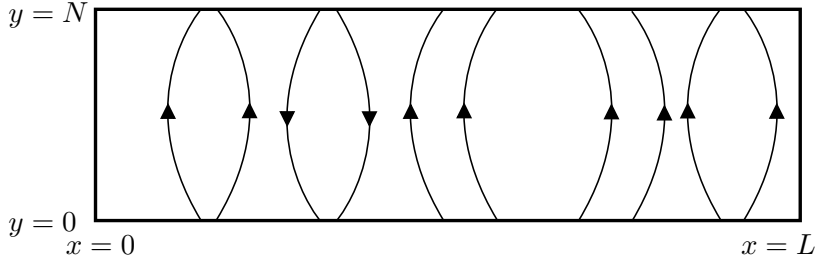}
\caption{Coulomb gas on the annulus. The boundaries $x=0$ and $x=L$ are identified, but there can be a difference of height $h(x+L,y)= 2p \pi  + h(x,y) , \quad p \in \Z
$ because the charge is not conserved along a boundary (see also figure $\ref{fig:TLbloborient}$). Here $p=3$.}
\label{Coulomb_annulus_m0}
\end{figure}

\begin{figure}[htbp]
\centering
\psfrag{alpha1}[l]{$x=0$}
\psfrag{alpha2}[l]{$x=L$}
\psfrag{beta1}[l]{$y=0$}
\psfrag{beta2}[l]{$y=N$}
\qquad \includegraphics[width=0.7\textwidth]{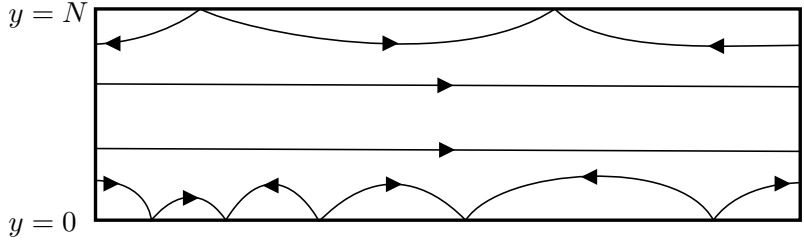}
\caption{The non-contractible loops in the Coulomb gas framework.
  Those touching the boundaries become an ensemble of half-loops
  between points on the boundary. Each non-contractible loop in the
  bulk contributes to the character $K_0$ with a weight $2$.}
\label{Coulomb_annulus_j}
\end{figure}

In addition, we must treat the non-contractible loops. Note that a
non-contractible loop which touch the boundary is no longer a loop in
our prescription for the Coulomb gas, because it is broken in several
half-loops on the boundary. However, the non-contractible loops which
remain in the bulk must be counted correctly (see
Fig.~$\ref{Coulomb_annulus_j}$). Remember that we do not want to
compute the full partition function here, but only the character $K_0$
corresponding to the representation of $2BTL$ without string $\V_0$.
Consider the following instructive example: we want to compute the
trace over the module $\V_0$ of the following element of $2BTL$
$$
e_1 b_1=\begin{pspicture}(0.,0.2)(1.4,0.7)
	\psline(0.9,0.)(0.9,0.6)
	\psline(1.3,0.)(1.3,0.6)
	\pscircle*(0.15,0.15){0.1}
	\psellipticarc{-}(0.3,0.6)(0.2,0.25){180}{0}
	\psellipticarc{-}(0.3,0.)(0.2,0.25){0}{180}
\end{pspicture}
$$
Only $4$ states in $\V_0$ contribute to this trace
$$
\begin{pspicture}(0.,0.)(1.3,0.7)
	\psellipticarc{-}(0.3,0.5)(0.2,0.5){180}{0}
	\psellipticarc{-}(0.9,0.5)(0.2,0.5){180}{0}
\end{pspicture}  \qquad \begin{pspicture}(0.,0.)(1.3,0.7)
	\pscircle*(0.75,0.25){0.1}
	\psellipticarc{-}(0.3,0.5)(0.2,0.5){180}{0}
	\psellipticarc{-}(0.9,0.5)(0.2,0.5){180}{0}
\end{pspicture} \qquad \begin{pspicture}(0.,0.)(1.3,0.7)
	\pscircle(1.05,0.25){0.1}
	\psellipticarc{-}(0.3,0.5)(0.2,0.5){180}{0}
	\psellipticarc{-}(0.9,0.5)(0.2,0.5){180}{0}
\end{pspicture} \qquad \begin{pspicture}(0.,0.)(1.3,0.7)
	\pscircle*(0.75,0.25){0.1}
	\pscircle(1.05,0.25){0.1}
	\psellipticarc{-}(0.3,0.5)(0.2,0.5){180}{0}
	\psellipticarc{-}(0.9,0.5)(0.2,0.5){180}{0}
\end{pspicture}.
$$
It should be clear why there are exactly $4=2^2$ states contributing
to the trace. The top of the diagram corresponding to our element $e_1
b_1$ puts strong constraints on these states. More precisely, it gives
all the information about the part which is disconnected from the
bottom of the diagram \cite{JFR}. Then if there are $2j$ lines (not
blobbed, as in our example) going from the bottom to the top of the
diagram, the states that contribute to the trace are exactly those we
can form with half-loops, wathever their blob status is. In other
words, the number of these states is the dimension of the module
$\V_0$ on $2j$ strands, that is $2^{2j}$. A more complete study of
this is given in \cite{JS2}. This conclusion is sufficient for our
discussion: each non-contractible loop in the bulk contributes with a
weight $2$ to the character $K_0$.

\smallskip

This has an important consequence for our Coulomb gas construction :
since each non-contractible loop that we cross when we go from one
boundary to another is weighted by $2$, we do not have to put some
additional electric charge to correct their weight (unlike the case of
the infinite cylinder that we discussed above).

\begin{figure}[htbp]
\centering
a.\includegraphics[width=0.45\textwidth]{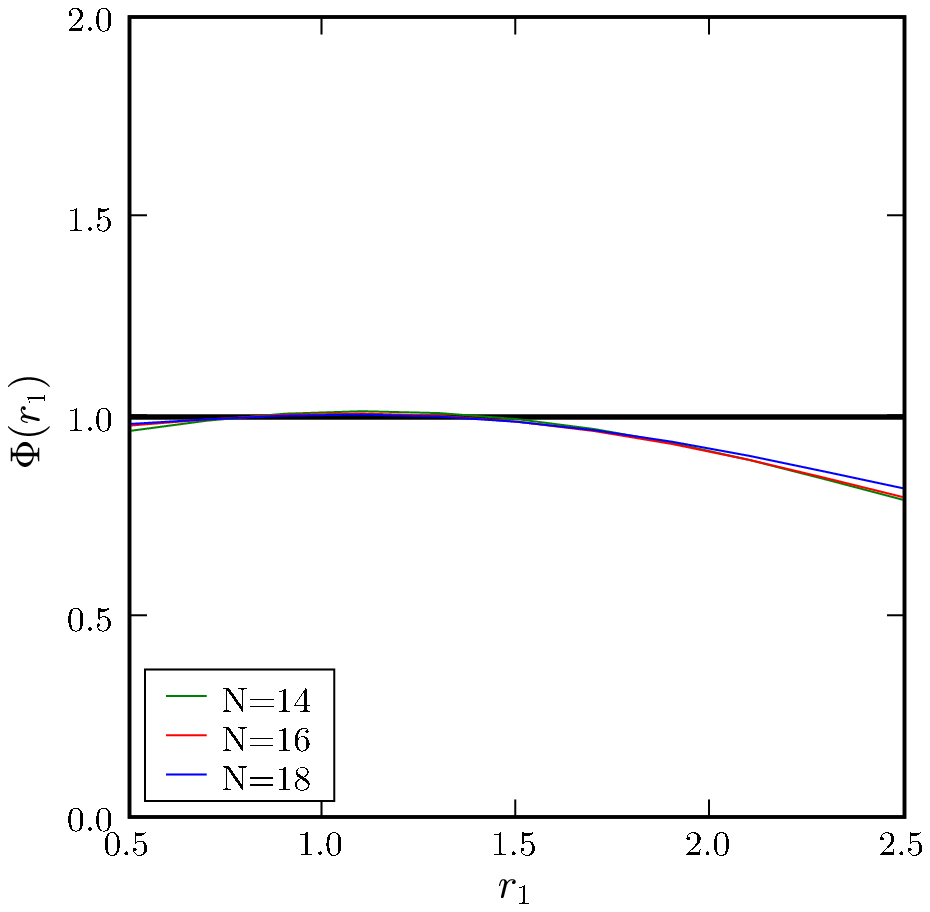} \quad b.\includegraphics[width=0.45\textwidth]{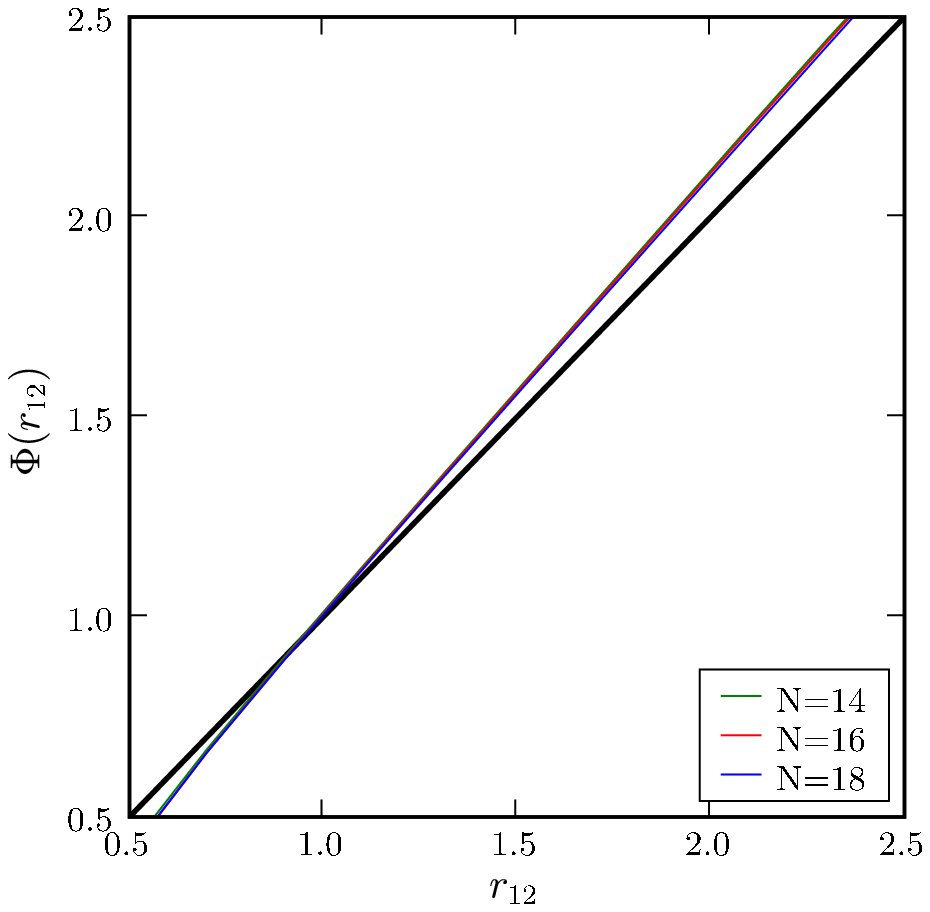}
\caption{Numerical results: we compute the largest eigenvalue of the
  transfer matrix, then exctract the leading exponent $h$ from the
  finite-size corrections. Here we plot the quantity $\Phi$ related to
  the exponent by $h=\frac{\Phi^2-1}{4m(m+1)}$ versus $r_1$ (a) and
  $r_{12}$ (b). Although the precision obtained here is not very
  satisfying, our conclusion is that $\Phi$ (and hence $h$) does not
  depend at all on $r_1$ and $r_2$ (a), but we rather have
  $\Phi=r_{12}$ (b).}
\label{fig:plot}
\end{figure}

Then, so far, we are able to count correctly the lines going from one
boundary to another, and the non-contractible loops in the bulk. The
next question is of course: what do we do with the vertices which
conserve the charge, which should be given weights proportional to
$e^{i r_1 \gamma}$, $e^{i r_2 \gamma}$, etc.? Our guess here is that
they do not contribute to the universal part of the character, so
$K_0$ does not depend at all on $r_1$ and $r_2$. The reason for this
is that the part involving $r_1$ (resp.\ $r_2$) in $b_1$ (resp.\
$b_2$) is diagonal, so it can be viewed equivalently as a field living
on the boundary. We expect any such boundary field to flow towards a
fixed boundary condition under RG, which should not depend on $r_1$
(or $r_2$). We have checked that conjecture numerically, by
transfer-matrix diagonalization and extraction of the finite-size
corrections. On the infinite strip of width $N$, the leading exponent
$h$ is related to finite-size corrections to the free energy per area
unit through the well-known relation
\begin{equation}
\label{FSC}
f_{N}=f_{\mathrm{bulk}} + \frac{f_{\mathrm{boundary}}}{N} + \frac{\pi h- \pi c/24}{N^2} + \mathcal{O}\left( \frac{1}{N^3}\right).
\end{equation}
We have computed $f_N$ for sizes up to $N=18$, then extracted the leading exponent $h$ using ($\ref{FSC}$) up to order $N^{-4}$. Although we do not reach a very satisfying precision, our numerical results are compatible with the conjecture that $h$ does only depend on $r_{12}$ (see Fig.~$\ref{fig:plot}$).

\smallskip

Now we are ready to compute the character $K_0$ itself. Let us
decompose $h(x,y)$ as
$$
h(x,y)=2p \pi + \tilde{h}(x,y)
$$
where $\tilde{h}(x+L,y)=\tilde{h}(x,y)$ and $\partial_y h (x,y=0) =
\partial_y h (x,y=N)=0$. The integration over $\tilde{h}$ gives the
usual $Z_0=q^{-1/24}/P(q)$. Then we are left with the contribution of
the height difference $2p\pi$, counted with a weight $e^{i p
  r_{12}\gamma}$ as explained above
$$
K_0 \propto Z_0 \sum_{p \in \Z} e^{i p r_{12} \gamma} e^{-(g/4\pi) p^2 (2 \pi/l)^2 (N L)} = Z_0 \sum_{p \in \Z} e^{i p r_{12} \gamma} e^{-(\pi g /\tau) p^2}
$$
where $\tau=L/N$. Now we want the expression of some Virasoro
character, so we have to work with $q=e^{-\pi \tau}$, not
$e^{-2\pi/\tau}$. We perform the Fourier transform using the Poisson
formula $\sum_p \rightarrow \sum_n \int dp e^{-2\pi i n p}$. The sum
becomes
\begin{eqnarray}
\nonumber \sum &=& \sum_n \int dp e^{-(\pi g/\tau) p^2 + i p (r_{12} \gamma - 2\pi n)} \\
\nonumber &=& \left( \tau/ g \right)^{1/2} \sum_n e^{-(\pi\tau/4 g) \left(r_{12} \gamma/\pi - 2 n\right)^2} \\
&=& \left( \tau/ g \right)^{1/2} \sum_n  q^{h_{r_{12}-2n,r_{12}}-(c-1)/24}.
\label{Poisson}
\end{eqnarray}
Normalizing the final expression such that the contribution of the
identity operator ($r_{12}=1$) without its descendents is just
$q^{-c/24}$, we end up with
\begin{equation}
K_0 = \frac{q^{-c/24}}{P(q)} \sum_{n \in \Z} q^{h_{r_{12}-2n,r_{12}}}.
\label{K_0}
\end{equation}
Note that, although this character depends only on $r_{12}$, it is not
true that it does not depend on the loop weights $n_1$ and $n_2$,
because all these parameters are linked by ($\ref{param:nb}$). Thus
the character $K_0$ is a function of $n$, $n_1$, $n_2$ and $n_{12}$ as
expected. Note also that, because of the parametrization
($\ref{param:nb}$), we expect that $K_0$ is invariant under
\begin{equation}
r_{12} \rightarrow -r_{12}
\end{equation}
and
\begin{equation}
r_{12} \rightarrow r_{12}+2 \pi/\gamma =r_{12}+2 (m+1)
\end{equation}
which is indeed the case for ($\ref{K_0}$), because of the symmetries
of Kac's formula ($\ref{Kac}$).


\subsection{Relation with the open XXZ spin chain\footnote{This
    digression can be skipped at the first lecture.}}

The fact that the representation $\V_0$ of $2BTL_N$ is exactly of size
$2^N$ and the oriented loop framework we developed above both suggest
that there is some link with the celebrated spin $1/2$ XXZ chain, with
appropriate boundary conditions. We would like to develop a bit this
subject in the following section. In fact, the equivalence between the
representation $\V_0$ presented above and the so-called spin chain
representation of $2BTL_N$ was proved in \cite{deGier2BTL}.

\smallskip

It is well-known that the Temperley-Lieb generators $e_i$ can be
interpreted as a local Hamiltonian density, that is we can construct a
simple Hamiltonian (here with the blob operators)
\begin{equation}
\mathcal{H} = -\lambda_1 b_1 - \lambda_2 b_2 -\sum_{i=1}^{N-1} e_i
\label{HamDensity}
\end{equation}
where $\lambda_1$ and $\lambda_2$ are two (so far unknown) constants.

\begin{eqnarray*}
e_i &=& -\frac{1}{2} \left( \sigma_{i}^x\sigma_{i}^x+ \sigma_{i}^y\sigma_{i}^y+ \cos \gamma \; \sigma_{i}^z\sigma_{i}^z \right) + i \frac{\sin \gamma}{2} \left(\sigma_{i}^z-\sigma_{i+1}^z \right)+\frac{\cos \gamma}{2} \\
b_1 &=& -\frac{1}{2 \sin r_1 \gamma} \left( \sin s_1 \gamma \; \sigma_{1}^z + \cos s_1 \gamma \; \sigma_{1}^z+ i \cos r_1 \gamma \; \sigma_{1}^z \right) + \frac{1}{2} \\ 
b_2 &=& \frac{1}{2 \sin r_2 \gamma} \left( \sin s_2 \gamma \; \sigma_{1}^z + \cos s_2 \gamma \; \sigma_{1}^z+ i \cos r_2 \gamma \; \sigma_{1}^z \right) + \frac{1}{2} \\
\end{eqnarray*}
with
\begin{equation}
r_{12}=s_2-s_1.
\label{s1s2r12}
\end{equation}
If we parametrize
\begin{equation}
\lambda_1 = \frac{\sin \gamma \; \sin r_1 \gamma}{\sin \phi_1 \; \sin (r_1 \gamma +\phi_1)} \qquad \lambda_2 = \frac{\sin \gamma \; \sin r_2 \gamma}{\sin \phi_2 \; \sin (r_2 \gamma +\phi_2)}.
\end{equation}
then our Hamiltonian is, up to an irrelevant additive constant
\begin{eqnarray}
\label{XXZham}
\mathcal{H} &=& \frac{1}{2} \left\{ \sum_{i=1}^{N-1} \left( \sigma_{i}^x\sigma_{i}^x+ \sigma_{i}^y\sigma_{i}^y+ \cos \gamma \; \sigma_{i}^z\sigma_{i}^z \right) \right. \\
\nonumber &+& \sin \gamma\left[ \frac{1 }{\sin \phi_1 \; \sin (r_1 \gamma + \phi_1)} \left(\sin s_1 \gamma \; \sigma_1^x + \cos s_1 \gamma \; \sigma_1^y \right) + i \rm{cotan} \phi_1 \; \rm{cotan} (r_1 \gamma+\phi_1) \; \sigma_1^z \right] \\
\nonumber &-& \left. \sin \gamma\left[ \frac{1 }{\sin \phi_2 \; \sin (r_2 \gamma + \phi_2)} \left(\sin s_2 \gamma \; \sigma_N^x + \cos s_2 \gamma \; \sigma_N^y \right) + i \rm{cotan} \phi_2 \; \rm{cotan} (r_2 \gamma+\phi_2) \; \sigma_N^z \right] \right\}
\end{eqnarray}
Note that this is a Hamiltonian for the XXZ chain with non-diagonal
boundary terms. This kind of Hamiltonian has been studied in great
detail over the recent years
\cite{deGierPyatov,Nepomechie,Nepomechie2}. Note also that this
Hamiltonian is not hermitian, which was already the case for closed
boundaries with an $SU(2)_q$ symmetry \cite{PasquierSaleur}.

Our derivation of the character $K_0$ ($\ref{K_0}$) applies directly
to the Hamiltonian ($\ref{XXZham}$), so we make the following
conjecture about the spectrum of this spin chain. The universal part
of the spectrum of $\mathcal{H}$ does not depend on $\lambda_1$ and
$\lambda_2$ when these are positive real numbers. This has been
discussed in some detail in \cite{JS1} in the case of one boundary,
and we expect this to be true also in the present case. Then the
spectrum should depend neither on the parameters $\phi_1$, $\phi_2$,
nor on $r_1$ and $r_2$. The only relevant parameter is the difference
$s_2-s_1$, which is related to the weights of the loops touching both
boundaries in the loop model via ($\ref{s1s2r12}$) and
($\ref{param:nb}$). The spectrum of the XXZ Hamiltonian
($\ref{XXZham}$) is then given by
\begin{equation}
E_n=\frac{\pi v_F}{N} \left( h_{r_{12-2n,r_{12}}}-c/24\right).
\end{equation}
where $v_F = \frac{\pi \sin \gamma}{\gamma}$ is the "Fermi velocity".


\section{The two-boundary partition function}

\subsection{One-boundary case}

In \cite{JS1}, it has been conjectured that the partition function on
the annulus with one free boundary condition and one blob is
\begin{equation}
Z_{1B} = \frac{q^{-c/24}}{P(q)} \left\{ \sum_{j \geq 0} \frac{\sin (u_1+2j) \chi}{\sin u_1 \chi} q^{h_{r_1,r_1+2j}} + \sum_{j \geq 1} \frac{\sin (-u_1+2j)}{\sin -u_1 \gamma} q^{h_{-r_1,-r_1+2j}} \right\}
\label{Z1B}
\end{equation}
where we recognize again some polynomials in $l=2 \cos\chi$ and $l_1=
\frac{\sin (u_1+1) \chi}{\sin u_1 \chi}$. This partition function
hence has the structure we have detailed in the previous section, but
on the One-Boundary Temperley-Lieb algebra $1BTL$. So far, we have
failed to provide some Coulomb gas arguments to derive the exponents
$h_{r,r}$. Strong numerical evidence has been given in $\cite{JS1}$
for general $r_1$, and exact results have been obtained from Bethe
ansatz when $r_1$ is an integer \cite{Nichols}. Many of these results
have been rederived since by Kostov using 2d quantum gravity
techniques \cite{Kostov,Bourgine}. Note the consistency with our computation of
$K_0$ from the previous section : the leading exponent we expect from
($\ref{K_0}$) is $h_{r_{12},r_{12}}$. The one-boundary case should be
recovered from $n_2=n$ and $n_{12}=n_1$, that is $r_2=1$ and
$r_{12}=r_1$. We see that $h_{r_1,r_1}$ is indeed the leading exponent
appearing in ($\ref{Z1B}$). A more precise analysis of the relation
between the character $K_0$ for two boundaries and the character
$q^{h_{r_1,r_1}}/P(q)$ in ($\ref{Z1B}$) also exists, although it
recquires more representation theory for the algebra $2BTL$. We will
report on this in \cite{DJS}.


\subsection{Boundary states and the partition function}

\begin{figure}[htbp]
\centering
\psfrag{qqtilde}[c]{$q \rightarrow \q$}
\psfrag{evol1}[l]{$L_0-\frac{c}{24}$}
\psfrag{evol2}[l]{$L_0+\bar{L}_0-\frac{c}{12}$}
\psfrag{modular}[l]{Modular}
\psfrag{transform}[l]{transform}
\psfrag{beta1}[l]{$\ket{B_1}$}
\psfrag{beta2}[l]{$\ket{B_2}$}
\psfrag{alpha1}[l]{$B_1$}
\psfrag{alpha2}[l]{$B_2$}
\includegraphics[width=0.7\textwidth]{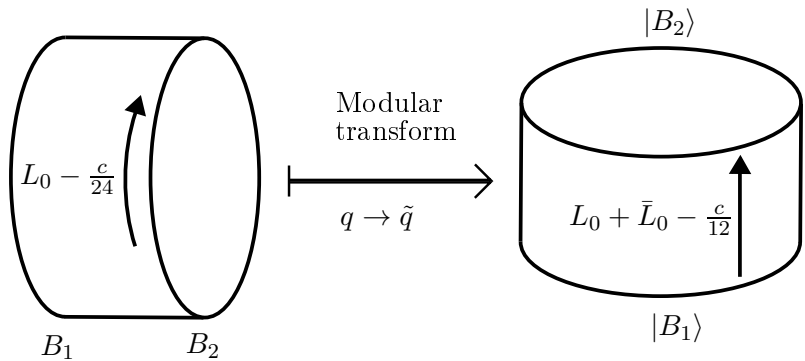}
\caption{Modular transform of the partition function. This corresponds
  to the ``open'' or ``closed string channel'' respectively. We have
  $Z_{B_1 B_2}= \bra{B_1} \q^{L_0+\bar{L}_0-c/12}\ket{B_2}.$}
\label{ModularTransform}
\end{figure}

Now we turn to the computation which is the core of this paper, and we
determine completely the partition function of our two-boundary loop
model in the most general case. The main idea of this computation
follows the work of Cardy on minimal theories
\cite{CardyBCFTVerlinde}. We start from the one-boundary partition
function $Z_{1B}$, and compute its modular transform. The result is
then interpreted as a scalar product between an initial boundary state
$\ket{B_1}$ and the final state $\ket{\mathrm{free}}$, with an
evolution operator $\q^{L_0 + \bar{L}_0 -c/12}=e^{-2\pi N/L (L_0 +
  \bar{L}_0 -c/12)}$ inserted (see Fig.~$\ref{ModularTransform}$). Then
we argue that this result together with the knowledge of the sector
without strings is sufficient to guess the partition function of the
form $\bra{B_2} \q^{L_0 + \bar{L}_0 -c/12} \ket{B_1}$. We conclude by
computing the modular transform back, and get the general partition
function in the form ($\ref{FormSuperGuess}$).

\paragraph{Modular transform of the one-boundary partition function:}
We start from ($\ref{Z1B}$) and use again the Poisson formula $\sum_j
\rightarrow \sum_p \int dj e^{i 2\pi j p}$, exactly as in
($\ref{Poisson}$).
\begin{equation}
Z_{1B} = (2 g)^{-1/2} \frac{\q^{-c/12}}{P(\q^2)} \sum_{p \in \Z} \frac{\sin \left( u_1 \chi + r_1 (\gamma/ g) ( p  + \chi/\pi) \right)}{\sin u_1 \chi} \q^{2 (1/4g) \left[ (\chi/\pi+p)^2 -(\gamma/\pi)^2 \right]}.
\end{equation}

\paragraph{What with loops touching both boundaries?}
Something special must happen in the sector without strings, because
of the loops touching both boundaries. The one-boundary partition
function may be seen as a very special case of the two-boundary one,
when $n_2=n$ and $n_{12}=n_1$. Thus, in the one-boundary partition
function, the character $K_0$ given by ($\ref{K_0}$) is present, with
the special value $r_{12}=r_1$. However, for a generic value of
$r_{12}$, the exponents $h_{r_{12}-2n,r_{12}}$ have no such special
value. On the other hand, we expect all the exponents in the string
sectors to be completely independent of $r_{12}$, since they cannot
depend on the weight of loops touching both boundaries. Hence, in this
respect, the sector without strings decouples from all the other
sectors. In particular, the formalism shown in
Fig.~$\ref{ModularTransform}$ should apply if we simply cancel the
contribution $K_0$ coming from the sector without strings. At the end
of the computation, because of the form of the partition function
($\ref{FormSuperGuess}$), it will be sufficient to add $K_0$ with
$r_{12}$ giving the correct weight $n_{12}$ to the loops touching both
boundaries, see ($\ref{param:nb}$). We have then
\begin{eqnarray}
\nonumber Z_{1B} - K_0(r_{12}=r_1) &=& (2 g)^{-1/2} \frac{\q^{-c/12}}{P(\q^2)} \sum_{p \in \Z} \frac{\sin \left( u_1 \chi + r_1 (\gamma/\pi g) ( p \pi + \chi) \right)}{\sin u_1 \chi} \q^{2 (1/4g) \left[ (\chi/\pi+p)^2 -(\gamma/\pi)^2 \right]} \\
\nonumber && - (2g)^{-1/2} \frac{\q^{-c/12}}{P(\q^2)} \sum_{p \in \Z} \cos\left( r_1 p \gamma \right) \q^{2 (g/4) \left[ p^2 - (\gamma/\pi g)^2 \right]}\\
&\equiv& \bra{{\rm free}} \q^{L_0 + \bar{L_0}-c/12} \ket{B(u_1,r_1)}.
\label{Scal1B}
\end{eqnarray}

\paragraph{Boundary states:} Recall that the free boundary condition
on the boundary $2$ actually corresponds to $u_2=r_2=1$. What we want
to do now is to identify the terms of
\begin{equation}
\bra{B(u_2,r_2)} \q^{L_0 +\bar{L}_0 -c/12} \ket{B(u_1,r_1)} = \frac{\q^{-c/12}}{P(\q^2)} \sum_{h_\alpha} \left< B(u_2,r_2) | h_\alpha \right>  \left< h_\alpha | B(u_1,r_1) \right> \q^{2 h_\alpha}
\label{ScalStrings}
\end{equation}
where the sum runs over all the primary exponents appearing in
($\ref{Scal1B}$), and the states $\ket{h_\alpha}$ satisfy $L_0
\ket{h_{\alpha}}= \bar{L}_0 \ket{h_\alpha} = h_\alpha \ket{h_\alpha}$.
We have to distinguish the two sets of exponents appearing in
($\ref{Scal1B}$).

\begin{itemize}
\item $h_\alpha = 1/4g \left[ (\chi/\pi+p)^2 -(\gamma/\pi)^2 \right]$:
Eq.~($\ref{Scal1B}$) gives
$$\left< B(1,1) | h_\alpha \right> \left< h_\alpha | B(u_1,r_1) \right>= \left(2g\right)^{-1/2}\frac{\sin \left( u_1 \chi +r_1 (\gamma/\pi g) (p \pi+\chi) \right)}{\sin u_1 \chi},$$
so we can guess that the straightforward generalization holds
\begin{eqnarray}
&&\left< B(u_2,r_2) | h_\alpha \right> \left< h_\alpha | B(u_1,r_1) \right>  \nonumber \\
&=& \left(2g\right)^{-1/2}\frac{\sin \chi}{\sin u_1 \chi \sin u_2 \chi} \frac{ \sin \left( u_1 \chi +r_1 (\gamma/\pi g) (p \pi+\chi) \right) \sin \left( u_2 \chi +r_2 (\gamma/\pi g) (p \pi+\chi) \right)}{\sin  \left( \chi + (\gamma/\pi g) (p \pi + \chi) \right)}
\label{guess1}
\end{eqnarray}

\item $h_\alpha = g/4 \left[p^2 -(\gamma/\pi g)^2 \right]$:
This time Eq.~($\ref{Scal1B}$) seems to give simply
$$\left< B(1,1) | h_\alpha \right> \left< h_\alpha | B(u_1,r_1) \right>= -\left(2g\right)^{-1/2}\cos \left( r_1 p \gamma \right),$$
a result which is independent of $u_1$. This actually would lead to
absurd conclusions. Indeed, we work in the string sectors, which all
give non-contractible loops, so we expect all the terms to be affected
somehow by the weights $l$, $l_1$ and $l_2$. The only terms which are
completely independent of these weights appear in the sector without
string, and we have cancelled this contribution. This contradiction
comes from the fact that $h_\alpha$ is even in $p$, so when we take
the sum over all the exponents, only the even part of $\left< B(1,1) |
  h_\alpha \right> \left< h_\alpha | B(u_1,r_1) \right>$ remains.
Inspired by the form of the coefficients we have already encountered,
we can try the simple but non-trivial inclusion of the following odd
term in $p$
$$\left< B(1,1) | h_\alpha \right> \left< h_\alpha | B(u_1,r_1) \right>= -\left(2g\right)^{-1/2}\frac{\sin (u_1 \chi) \cos \left( r_1 p \gamma \right)+ \cos (u_1 \chi) \sin \left( r_1 p \gamma \right)}{\sin u_1 \chi},$$
leading to the generalization
\begin{eqnarray}
&&\left< B(u_2,r_2) | h_\alpha \right> \left< h_\alpha | B(u_1,r_1) \right>  \nonumber \\
&&= -\left(2g\right)^{-1/2}\frac{\sin \chi}{\sin u_1 \chi \sin u_2 \chi} \frac{ \sin \left( u_1 \chi + p r_1 \gamma\right) \sin \left( u_2 \chi +p r_2\gamma \right)}{\sin \left( \chi + p \gamma \right)}
\label{guess2}
\end{eqnarray}

\end{itemize}

\paragraph{Modular transform:}
Although the (partly guessed) relations ($\ref{guess1}$) and
($\ref{guess2}$) seem quite complicated, they lead to quite a nice
formula when we go back to the ``open string channel'' (see
Fig.~$\ref{ModularTransform}$). To see this, we need once again to
perform a modular transform. The two sums appearing in
($\ref{ScalStrings}$) are now
\begin{eqnarray}
Z_+ = - (2g)^{-1/2} \frac{\q^{-c/12}}{P(\q^2)} \sum_{p \in \Z} \frac{\sin \chi}{\sin u_1 \chi \sin u_2 \chi} \frac{ \sin \left( u_1 \chi + p r_1 \gamma\right) \sin \left( u_2 \chi +p r_2\gamma \right)}{\sin \left( \chi + p \gamma \right)} \q^{2(g/4)\left[p^2-(\gamma/\pi g)^2\right]}
\end{eqnarray}
and
\begin{eqnarray}
\nonumber Z_- &=& (2g)^{-1/2} \frac{\q^{-c/12}}{P(\q^2)}  \sum_{p \in \Z}  \frac{\sin \chi}{\sin u_1 \chi \sin u_2 \chi}\\
&\times& \frac{ \sin \left( u_1 \chi +r_1 (\gamma/\pi g) (p \pi+\chi) \right) \sin \left( u_2 \chi +r_2 (\gamma/ \pi g) (p \pi+\chi) \right)}{\sin  \left( \chi + (\gamma/\pi g) (p \pi + \chi) \right)} \q^{2(1/4g)\left[(\chi/\pi+p)^2-(\gamma/\pi)^2\right]^2}.
\end{eqnarray}
We can compute the modular transform of each part independently, and
add the contributions in the end. Let us begin with $Z_+$. The product
can be decomposed as
\begin{eqnarray*}
&-& \frac{\sin \left(u_1 \chi + p r_1 \gamma\right) \sin \left( u_2+p r_2 \gamma \right)}{\sin \left( \chi + p\gamma \right)} \\
&=&\frac{1}{2} \rm{Im} \; \sum_{\epsilon_{1,2}=\pm 1}\sum_{j \geq 1} \epsilon_1 \epsilon_2 e^{i \left[ (\epsilon_1 u_1 +\epsilon_2 u_2 -1 +2j)\chi +p \pi g ((\epsilon_1 r_1+\epsilon_2 r_2-1)(\gamma/\pi g)-2j) \right]}.
\end{eqnarray*}
$Z_+$ is then of the form
$$
Z_+ = \sum_{\epsilon_{1,2}=\pm 1} \sum_{j \geq 1} Z_+(j,\epsilon_{1,2})
$$
and $Z_+(j,\epsilon_{1,2})$ is a sum over $p \in \Z$. Let us write
$R=\epsilon_1 r_1 + \epsilon_2 r_2-1$ and $U=\epsilon_1 u_1 +
\epsilon_2 u_2-1$. We use the Poisson formula $\sum_{p \in \Z}
\rightarrow \sum_{n \in \Z} \int dp e^{i2\pi p n}$ to compute the
modular transform of $Z_+(j,\epsilon_{1,2})$. Note that we also use
$\q^{-1/12}/P(\q^2)=(\tau /2)^{-1/2}q^{-1/24}/P(q)$ as usual. The sum
appearing in the computation is
\begin{eqnarray*}
&& \sum_{p \in \Z} e^{-(\pi/\tau) g p^2} e^{i p \pi g (R \gamma/\pi g-2j)} \\
&=& \sum_{n \in \Z} \int dp \; e^{i 2\pi p n} e^{-(\pi/\tau)g p^2} e^{i p \pi g (R \gamma/\pi g-2j)}\\
&=& (\tau g)^{1/2} \sum_{n \in \Z} q^{h_{R-2n, R+2j}-(c-1)/24}
\end{eqnarray*}
Putting all things together, we get
\begin{equation}
  Z_+ = \frac{1}{2} \frac{q^{-c/24}}{P(q)} \sum_{\epsilon_{1,2}=\pm 1} \sum_{j \geq 1} \sum_{n \in \Z} \frac{\sin (\epsilon_1 u_1+\epsilon_2 u_2 -1 +2j)\chi \sin \chi }{\sin \epsilon_1 u_1 \chi \sin \epsilon_2 u_2 \chi} q^{h_{\epsilon_1 r_1 +\epsilon_2 r_2 -1-2n ,\epsilon_1 r_1 +\epsilon_2 r_2 -1+2j}}. 
\label{Zplus}
\end{equation}
Now consider the case of $Z_-$. The computation is quite similar.
First we have to decompose the sinus product. Let $x=(\gamma/g)
(p+\chi/\pi)$, then
\begin{eqnarray*}
&&\frac{\sin(r_1 x +u_1) \sin(r_2 x +u_2 \chi)}{\sin(x+\chi)} \\
&=& \frac{1}{2} \rm{Im} \; \sum_{\epsilon_{1,2}=\pm 1}\sum_{n \geq 0} \epsilon_1 \epsilon_2 e^{i \left[ (\epsilon_1 r_1 + \epsilon_2 r_2 -1-2n)x+(\epsilon_1 u_1+\epsilon_2 u_2 -1-2n)\chi \right]}
\end{eqnarray*} 
Then we use Poisson's formula $\sum_{p \in \Z} \rightarrow \sum_{j \in \Z} \int dp e^{i2\pi j p}$ to get in the end
$$
Z_-=\frac{1}{2} \frac{q^{-c/24}}{P(q)} \sum_{\epsilon_{1,2}=\pm 1} \sum_{j \in \Z} \sum_{n \geq 0} \frac{\sin (\epsilon_1 u_1+\epsilon_2 u_2 -1 +2j)\chi \sin \chi }{\sin \epsilon_1 u_1 \chi \sin \epsilon_2 u_2 \chi} q^{h_{\epsilon_1 r_1 +\epsilon_2 r_2 -1-2n ,\epsilon_1 r_1 +\epsilon_2 r_2 -1+2j}}
$$
and after some relabelling of the indices, we have
\begin{eqnarray}
\nonumber Z_-&=&\frac{1}{2} \frac{q^{-c/24}}{P(q)} \sum_{\epsilon_{1,2}=\pm 1} \sum_{j \geq 1} \frac{\sin (\epsilon_1 u_1+\epsilon_2 u_2 -1 +2j)\chi \sin \chi }{\sin \epsilon_1 u_1 \chi \sin \epsilon_2 u_2 \chi} \\
&& \times \left\{ \sum_{n \geq 0} q^{h_{\epsilon_1 r_1 +\epsilon_2 r_2 -1-2n ,\epsilon_1 r_1 +\epsilon_2 r_2 -1+2j}} - \sum_{n < 0} q^{h_{\epsilon_1 r_1 +\epsilon_2 r_2 -1-2n ,\epsilon_1 r_1 +\epsilon_2 r_2 -1+2j}} \right\}.
\label{Zminus}
\end{eqnarray}

\paragraph{The two-boundary partition function:} 
Adding the terms ($\ref{Zplus}$) and ($\ref{Zminus}$), we find that the total contribution of all the string sectors is
\begin{equation}
\frac{q^{-c/24}}{P(q)} \sum_{\epsilon_{1,2}=\pm 1} \sum_{j \geq 1} \sum_{n \geq 0} \frac{\sin (\epsilon_1 u_1+\epsilon_2 u_2 -1 +2j)\chi \sin \chi }{\sin \epsilon_1 u_1 \chi \sin \epsilon_2 u_2 \chi} q^{h_{\epsilon_1 r_1 +\epsilon_2 r_2 -1-2n ,\epsilon_1 r_1 +\epsilon_2 r_2 -1+2j}}. 
\end{equation}
If we now take into account the sector without strings and add its
conformal character $K_0$, we obtain the partition function
($\ref{SuperGuess}$) of our loop model, as claimed in the introduction
of this paper. This partition function has the form
($\ref{FormSuperGuess}$) as expected. We are now able to identify all
the conformal characters corresponding to the different sectors.
\begin{subequations}
\begin{eqnarray}
K_{2j}^{bb} &=& \frac{q^{-c/24}}{P(q)}\sum_{n \geq 0} q^{h_{r_1+r_2-1-2n,r_1+r_2-1+2j}} \\ 
K_{2j}^{bu} &=& \frac{q^{-c/24}}{P(q)}\sum_{n \geq 0} q^{h_{r_1-r_2-1-2n,r_1-r_2-1+2j}} \\ 
K_{2j}^{ub} &=& \frac{q^{-c/24}}{P(q)}\sum_{n \geq 0} q^{h_{-r_1+r_2-1-2n,-r_1+r_2-1+2j}} \\ 
K_{2j}^{uu} &=& \frac{q^{-c/24}}{P(q)}\sum_{n \geq 0} q^{h_{-r_1-r_2-1-2n,-r_1-r_2-1+2j}} 
\end{eqnarray}
\end{subequations}
Note that these characters are all related by the blobbed/unblobbed
transformation ($\ref{blobbed_unblobbed}$).

%
%

\section{Comparison with known results}
We would like to check our partition function against some known results from \cite{CardyBCFTVerlinde,CardyPercolationAnnulus,BauerSaleur}. 

\subsection{Critical percolation on the annulus}
As a first simple application of our result, we can turn to the
critical percolation problem on an annulus. Critical percolation
corresponds to $l=l_1=l_2=n=n_1=n_2=n_{12}=1$, and in that case the
partition function ($\ref{SuperGuess}$) is simply
\begin{equation}
Z=1.
\label{perco1}
\end{equation}
To see this, it is sufficient to note that our partition function
reduces to the one-boundary partition function ($\ref{Z1B}$) when
$l_2=n_2=n$ and $n_{12}=n_1$. The one-boundary case itself reduces to
the free/free case when $l_1=n_1=n$. Then for $l=n=1$, it is easy to
see that $Z=1$ using Euler's pentagonal identity. Now if we want to
compute, for example, the probability $P_{\mathrm{crossing}}$ that
there is at least one contractible percolation cluster going from one
boundary to the other, we have to vary the weight of loops touching
both boundaries $n_{12}$. Indeed, each percolation cluster is
encircled by exactly one loop, and each cluster touches a boundary if
and only if its surrounding loop touches it. Since we know that
$n_{12}$ does only appear through $r_{12}$ in the conformal character
$K_0$, we have
\begin{equation}
Z(n_{12})=1+ K_0(r_{12}) - K_0(r_{12}=1)
\label{perco2}
\end{equation}
and then
\begin{eqnarray}
\nonumber P_{\mathrm{crossing}}&=&1-Z(n_{12}=0)\\
\nonumber &=& K_0(r_{12}=1)-K_0(r_{12}=3) \\
&=& \frac{\sum_{k \in \Z} \left( q^{6k^2+k} + q^{6k^2+5k+1} -2 q^{6 k^2+3k+\frac{1}{3}} \right)}{\prod_{k \geq 1} \left(1-q^k \right)}
\end{eqnarray}
which agrees with \cite{CardyPercolationAnnulus}.

\subsection{Relation with $Q$-state Potts models}

\begin{figure}[htbp]
\centering
\includegraphics[width=0.6\textwidth]{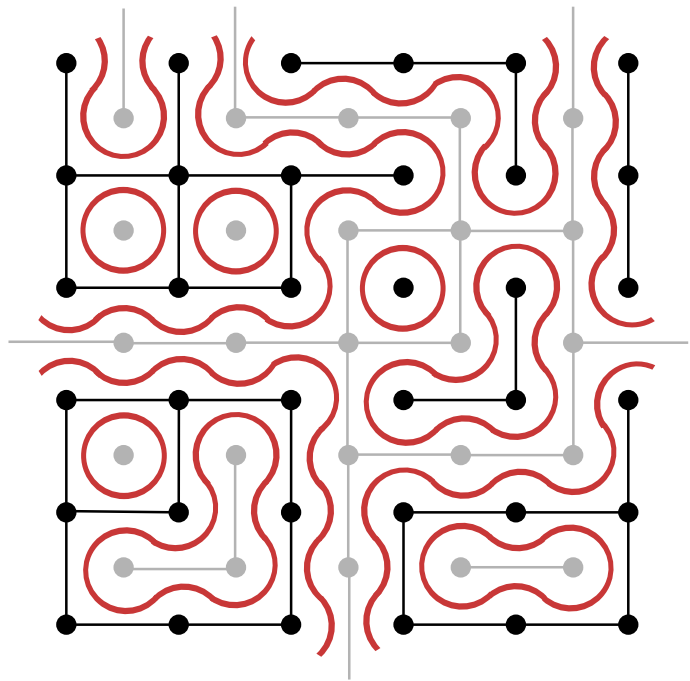}
\caption{Mapping from the Potts model to the loop model. The black
  structures are the FK clusters, while dual clusters are in grey.}
\label{Potts_loop}
\end{figure}

It is a well-known result that the $Q$-state Potts model can be
reformulated as a dense loop gas with the loop fugacity $n= \sqrt{Q}$.
Let us recall here how this can be achieved. The Potts model can be
defined on the square lattice, with a spin $\sigma_x \in \left\{ 1,
  \dots,Q \right\}$ living on each site. Only the neighbouring sites
interact, and the partition function is the sum over all the Potts
spin configurations
\begin{equation}
Z_{\mathrm{Potts}}= \sum_{\mathrm{Potts}} \prod_{<xx'>} \exp \left\{ K \delta(\sigma_x,\sigma_{x'}) \right\}
\label{PottsDef}
\end{equation}
and this is rewritten as
\begin{equation}
Z_{\mathrm{Potts}}= \sum_{\mathrm{Potts}} \prod_{<xx' >} \left( 1 +  \delta_{x,x'} v \right)
\label{PottsDefalt}
\end{equation}
with $v=e^K-1$. Then the most important step is to interpret
($\ref{PottsDefalt}$) as a random-cluster (Fortuin-Kasteleyn)
partition function. For a given Potts configuration, the FK clusters
live inside the Potts clusters.
\begin{equation}
Z_{\mathrm{Potts}}= \sum_{\mathrm{Potts}} \quad \sum_{\mathrm{FK \subset Potts}} v^{\# \mathrm{FK \; bonds}}
\label{PottsFK0}
\end{equation}
Taking the trace over the Potts configurations gives the Potts
partition function in its Fortuin-Kasteleyn representation
\begin{equation}
Z_{\mathrm{Potts}}= \sum_{\mathrm{FK}} v^{\# \mathrm{FK \; bonds}} Q^{\# \mathrm{FK \; clusters}}.
\label{PottsFK}
\end{equation}
In the FK representation, the mapping to the loop model is obvious:
one has to draw all the loops which encircle the FK clusters or the
clusters on the dual lattice (see Fig.~$\ref{Potts_loop}$). Now let $N$
be the number of loops, $C$ the number of clusters and $C^*$ the number
of dual clusters. Clearly, $N=C+C^*$. Moreover, Euler's formula gives
$C = C^* - \# \mathrm{FK \; bonds} + \# \mathrm{lattice \; vertices}$.
Then, up to an unimportant global factor, ($\ref{PottsFK}$) becomes
\begin{equation}
Z_{\mathrm{Potts}}= \sum_{\mathrm{Loop}} \left( \frac{v}{\sqrt{Q}} \right)^{\# \mathrm{FK \; bonds}} \sqrt{Q}^{N}.
\label{PottsLoop}
\end{equation}
It is well-known that the Potts model is critical when it is satisfies
the self-duality relation $v/\sqrt{Q}=1$. In that case,
($\ref{PottsLoop}$) is exactly the partition function of a loop gas
with fugacity $\sqrt{Q}$.

\smallskip

It is not difficult to generalize the previous discussion to the
boundary case. Let us assume that the Potts spins living on the
boundaries are restricted to some subsets $S_1$ and $S_2 \subset
\left\{1,\dots,Q\right\}$. Let
\begin{equation}
Q_1=\left| S_1 \right|  \qquad   Q_2=\left| S_2 \right|  \qquad   Q_{12}=\left| S_1 \cap S_2 \right|
\end{equation}
then taking again the trace over Potts configurations we get the
following relation instead of ($\ref{PottsFK}$)
\begin{equation}
Z_{\mathrm{Potts}}= \sum_{\mathrm{FK}} v^{\# \mathrm{FK \; bonds}} Q^{C} Q_1^{C_1} Q_2^{C_2} Q_{12}^{C_{12}}
\label{PottsFKbound}
\end{equation}
where $C$ is the number of bulk clusters, and $C_1$, $C_2$, $C_{12}$
are the number of FK clusters touching the boundary $1$, $2$, or both
of them. Introducing the number of loops of the same type $N_1$,
$N_2$, $N_{12}$, it is clear that $N_1=C_1$, $N_2=C_2$ and
$N_{12}=C_{12}$ since each boundary cluster is encircled by exactly
one boundary loop. For the bulk loops, this is different because each
one can encircle either a FK cluster or a dual cluster, so we still
have $N=C+C^*$. Now Euler's relation gives $C+C_1+C_2+C_{12}=C^* - \#
\mathrm{FK \; bonds} + \# \mathrm{lattice \; vertices}$. Up to a
global factor, ($\ref{PottsFKbound}$) becomes
\begin{equation}
Z_{\mathrm{Potts}}= \sum_{\mathrm{Loop}} \left(\frac{v}{\sqrt{Q}}\right)^{\# \mathrm{FK \; bonds}} \sqrt{Q}^N \left(\frac{Q_1}{\sqrt{Q}}\right)^{N_1} \left(\frac{Q_2}{\sqrt{Q}}\right)^{N_2} \left(\frac{Q_{12}}{\sqrt{Q}}\right)^{N_{12}}
\end{equation}
Again, we can impose the self-duality relation $v=\sqrt{Q}$ and then the identification of the loop weights is straightforward:
\begin{equation}
n=\sqrt{Q} \qquad n_1=\frac{Q_1}{\sqrt{Q}} \qquad n_2=\frac{Q_2}{\sqrt{Q}} \qquad n_{12}=\frac{Q_{12}}{\sqrt{Q}}
\label{Pottsloop_weights}
\end{equation}
At this point we have given the correct weight to all contractible
loops. In this article we are interested in a loop model on an
annulus, so we have to take care about the non-contractible loops.
This turns out to be non-trivial, and rather crucial if we want to
recover some known partition functions of the Potts model on the
annulus. The subtlety comes from the non-contractible FK clusters
which touch both boundaries, which must be restricted to the set $S_1
\cap S_2$. However, in the loop model these configurations are those
with exactly two non-contractible loop, each one touching one
boundary. Such configurations are counted with a weight $l_1 l_2 \neq
Q_{12}$. To solve this problem, we must identify the term coming with
the coefficient $l_1 l_2$ in the loop partition function
($\ref{SuperGuess}$), and give it the correct weight $Q_{12}$ to get
the Potts partition function.

\smallskip

Let $Z_{l_1 l_2}$ be this term in the loop partition function. To
identify this term, it is necessary to analyse carefully the
polynomials ($\ref{amplitudes}$). These can actually be written in
terms of the Chebyshev polynomials of the second kind $U_n(x)$, as
\cite{JS2}
\begin{equation}
 D_{2j}^{bb} = l_1 l_2 U_{2j-2}(l/2) - (l_1+l_2) U_{2j-3}(l/2) + U_{2j-4}(l/2)
\end{equation}
with similar expressions for the other polynomials $D_{2j}^{\alpha
  \beta}$, obtained by using the blobbed/unblobbed transformation
($\ref{blobbed_unblobbed}$), which maps $l_1$ on $l-l_1$ and/or $l_2$
on $l-l_2$. With those relations the identification of $Z_{l_1 l_2}$
is straightforward, noting that the constant coefficient of the
polynomial $U_{2n}$ is $(-1)^n$.
\begin{equation}
Z_{l_1 l_2} = \frac{q^{-c/24}}{P(q)} \sum_{j \geq 1} (-1)^{j-1} \left\{ K_{2j}^{bb}-K_{2j}^{ub}-K_{2j}^{bu}+K_{2j}^{uu} \right\}
\label{correction}
\end{equation}
Thus we have found the precise relation between our loop partition
function and the Potts one
\begin{equation}
Z_{\mathrm{Potts}}=Z_{\mathrm{loop}}+\left(Q_{12}-l_1 l_2 \right) Z_{l_1 l_2}
\label{PottsFinal}
\end{equation}
where all the loop weights are given by ($\ref{Pottsloop_weights}$)
for the contractible loops, and $l=n$, $l_1=n_1$, $l_2=n_2$ for
non-contractible ones. Of course we could have improved slightly the
mapping by distinguishing Potts clusters according to homotopy.
However, this will not be necessary to recover the known results about
the Potts model.

\subsubsection{Ising model}

We would like to use ($\ref{PottsFinal}$) to recover some results
about the Ising model on an annulus, which appeared in
\cite{CardyBCFTVerlinde,BauerSaleur}. Assume for example that the
Ising spins are fixed to $+$ on the first boundary and to $-$ on the
second one. This corresponds in our formalism to $Q=2$, $Q_1=Q_2=1$
and $Q_{12}=0$. Then all the parameters (see Table~$\ref{param:table}$
for the parametrizations) of the loop model are fixed: we have
$\gamma=\chi=\pi/4$, $u_1=r_1=u_2=r_2=2$, $r_{12}=3$.
Eq.~($\ref{SuperGuess}$) then gives
\begin{eqnarray}
\nonumber Z_{\mathrm{loop}} &=& \qc \sum_{n \in \Z} q^{h_{3-2n,3}} + \qc \sum_{j \geq 1} \frac{\sqrt{2}}{2} \left\{ \sin \left( 3+2j \right)\frac{\pi}{4} \sum_{n \geq 0} q^{h_{3-2n,3+2j}} \right. \\
 && \left. -2 \sin \left( -1+2j \right)\frac{\pi}{4} \sum_{n \geq 0} q^{h_{-1-2n,-1+2j}}  + \sin \left( -5+2j \right)\frac{\pi}{4} \sum_{n \geq 0} q^{h_{-5-2n,-5+2j}} \right\}
\end{eqnarray}
and adding the term ($\ref{correction}$) as in ($\ref{PottsFinal}$),
we get the Ising partition function
\begin{eqnarray}
\nonumber Z_{+/-} &=& \qc \sum_{n \in \Z} q^{h_{3-2n,3}} + \qc \sum_{j \geq 1} \frac{1}{2} \left\{ \left( \sqrt{2} \sin \left( 3+2j \right)\frac{\pi}{4} + (-1)^j \right)\sum_{n \geq 0} q^{h_{3-2n,3+2j}} \right. \\
\nonumber  &&  -2 \left( \sqrt{2} \sin \left( -1+2j \right)\frac{\pi}{4} +(-1)^j \right) \sum_{n \geq 0} q^{h_{-1-2n,-1+2j}} \\
 && \left. + \left( \sqrt{2} \sin \left( -5+2j \right)\frac{\pi}{4} +(-1)^j\right) \sum_{n \geq 0} q^{h_{-5-2n,-5+2j}} \right\}
\end{eqnarray}
Consider the second term between the brackets, which comes with a
factor $-2$. Consider this term twice, and once make the reindexation
$j \rightarrow j+2$, $n \rightarrow n-2$, and the second time $j
\rightarrow j-2$, $n \rightarrow n+2$. The first term thus obtained
cancels almost all the terms in the first sum between brackets, and
the second one almost all those of the third sum. Collecting what
remains after these cancellations, we have
\begin{eqnarray}
\nonumber Z_{+/-} &=& \qc \sum_{n \in \Z} q^{h_{3-2n,3}} \\
\nonumber && + \frac{1}{2} \qc \left\{ \sum_{j \geq 1} \left( \sqrt{2} \sin \left( 3+2j \right)\frac{\pi}{4} + (-1)^j \right)  \left( q^{h_{3,3+2j}} + q^{h_{1,3+2j}} \right)  \right. \\
\nonumber && - \left. \sum_{j \geq 3} \left( \sqrt{2} \sin \left( -5+2j \right)\frac{\pi}{4} + (-1)^j \right) \left( q^{h_{-1,-5+2j}} + q^{h_{-3,-5+2j}} \right)  \right. \\
&& \left. - 2 \sum_{n \geq 0}  \left( q^{h_{-5-2n,-3}} + q^{h_{-1-2n,3}} \right)  \right\}
\end{eqnarray}
Now we write $2j=8k+2, \; 8k+4, \; \ldots$ and then
\begin{eqnarray}
\nonumber Z_{+/-} &=& \qc \sum_{n \in \Z} q^{h_{3-2n,3}} \\
\nonumber && + \qc \sum_{k \geq 0} \left\{ q^{h_{3,3+8(k+1)}} + q^{h_{1,3+8(k+1)}} - q^{h_{3,3+2+8k}} -q^{h_{1,3+2+8k}}  \right. \\
\nonumber && - \left. q^{h_{-1,-5+8(k+1)}} - q^{h_{-3,-5+8(k+1)}} + q^{h_{-1,-5+2+8(k+1)}} + q^{h_{-3,-5+2+8(k+1)}} \right\} \\
&& -  \qc \sum_{n \geq 0} \left( q^{h_{-5-2n,-3}} +q^{h_{-1-2n,3}} \right)
\end{eqnarray}
Recall the Kac formula ($\ref{Kac}$) to see that $h_{-r,-s}=h_{r,s}$,
so all the terms combine to form the sums
\begin{equation}
Z_{+/-} = \qc \sum_{k \in \Z} \left\{ q^{h_{3,3+8k}} + q^{h_{1,3+8k}} - q^{h_{3,5+8k}} -q^{h_{1,5+8k}}  \right\}
\end{equation}
Again use the Kac formula and $m=3$ (recalling that $\gamma=\frac{\pi}{m+1}$)
to see that $h_{3,3+8k}=h_{3,5-8k}$ and then
\begin{equation}
Z_{+/-}=  \qc \sum_{k \in \Z} \left( q^{h_{1,3+8k}} - q^{h_{1,5+8k}} \right).
\end{equation}
Here we recognize the Rocha-Caridi formula, and we conclude that
\begin{equation}
Z_{+/-}=\chi_{1,3}
\end{equation}
as expected from \cite{CardyBCFTVerlinde}. We could have done the same
calculation for the spins fixed to $+$ on both boundaries. The
computation is exactly as the previous one, except that $Q_{12}=1$ so
$r_{12}=1$ this time. This would have led to
\begin{equation}
Z_{+/+}=\chi_{1,1}
\end{equation}
which is again a result of Cardy \cite{CardyBCFTVerlinde}. The other
boundary conditions, such as ${\rm free}/+$ for example, reduce to a
computation with the one-boundary partition function, which has been
studied in \cite{JS1}. Again all the results agree with those of
\cite{CardyBCFTVerlinde}.

\subsubsection{Three-states Potts model}

When $Q=3$ the Potts spins have three colours A,B,C. For example, we
can compute the partition function with all spins fixed to A or B with
equal probability on the first boundary, and to B or C on the second
one. We have then $Q_1=Q_2=2$, $Q_{12}=1$, so the parameters (see
Table~$\ref{param:table}$) of the loop model are $\gamma=\pi/6$,
$u_1=r_1=u_2=r_2=2$, $r_{12}=3$. The computation is exactly as in the
Ising case. Eq.~($\ref{PottsFinal}$) gives
\begin{eqnarray}
\nonumber Z_{AB/BC}&=& \qc \sum_{n \in \Z}  q^{h_{3-2n,3}} \\
\nonumber &&+ \frac{1}{3} \qc \sum_{j \geq 1} \left\{ \left( 2 \sin(3+2j) \frac{\pi}{6} +(-1)^j\right) \sum_{n \geq 0} q^{h_{3-2n,3+2j}} \right. \\
\nonumber &&-2\left( 2\sin (-1+2j) \frac{\pi}{6} +(-1)^j\right) \sum_{n \geq 0} q^{h_{-1-2n,-1+2j}} \\
&&+ \left.\left( 2 \sin(-5+2j) \frac{\pi}{6} +(-1)^j\right) \sum_{n\geq 0} q^{h_{-5-2n,-5+2j}} \right\}
\end{eqnarray} 
Once again we see that the double sums actually collapse to give
\begin{eqnarray}
\nonumber Z_{AB/BC}&=& \qc \sum_{n \in \Z}  q^{h_{3-2n,3}} \\
\nonumber &&+ \frac{1}{3} \qc \left\{ \sum_{j \geq 1} \left( 2 \sin(3+2j) \frac{\pi}{6} +(-1)^j\right) \left( q^{h_{3,3+2j}} +q^{h_{1,3+2j}} \right) \right. \\
\nonumber &&-\sum_{j \geq 3}\left( 2\sin (-5+2j) \frac{\pi}{6} +(-1)^j\right) \left( q^{h_{-1,-5+2j}} + q^{h_{-3,-5+2j}} \right) \\
\nonumber &&- 3 \left. \sum_{n \geq 0} \left( q^{h_{-1-2n,3}} + q^{h_{-5-2n,-3}} \right) \right\} \\
&=& \qc \sum_{k \in \Z} \left\{ q^{h_{1,3+12k}}-q^{h_{1,-3+12k}} +q^{h_{3,3+12k}} -q^{h_{3,-3+12k}} \right\}
\end{eqnarray} 
Using the Rocha-Caridi formula, we finally obtain
\begin{equation}
Z_{AB/BC}=\chi_{1,3}+\chi_{3,3}
\end{equation}
which agrees with \cite{CardyBCFTVerlinde}. All the results from this
reference concerning the Potts model can be deduced from our loop
partition function ($\ref{SuperGuess}$), with the relation
($\ref{PottsFinal}$).

\section{Refined crossing formulae for percolation on the annulus}

It should be obvious that the seven-parameter partition function
(\ref{SuperGuess}) harbours many more geometrical applications than the
known ones presented in the preceding section. As an illustration we
present here just one simple example.

Consider the continuum limit of critical percolation on an annulus of aspect
ratio $\tau=L/N$, and recall that $q={\rm e}^{-\pi \tau}$. Let $P_0$
be the probability that no cluster wraps the periodic direction, and
let $P^{\alpha \beta}_j$ be the probability that there are precisely
$j \ge 1$ wrapping clusters which are moreover constrained by the
values of the indices $\alpha,\beta$. When $\alpha=b$ (resp.\
$\alpha=u$) the leftmost cluster is constrained to touching
(resp.\ to not touching) the left rim; $\beta$ similarly constrains
the behaviour of the rightmost cluster.

Since $Z=1$ we have obviously
\begin{eqnarray}
 P_0 &=& Z \left( \chi=\frac{\pi}{2},u_1=1,u_2=1 \right) \nonumber \\
 \sum_{\alpha,\beta} P^{\alpha \beta}_j &=&
   \left. \frac{1}{(2j)!}
          \left( \frac{\partial_\chi}{\partial_l \chi} \right)^{2j}
          Z(u_1=1,u_2=1) \right|_{\chi=\frac{\pi}{2}} \nonumber \\
 P^{bb}_j &=&
   \left. \frac{1}{(2j-2)!}
          \left( \frac{\partial_\chi}{\partial_l \chi} \right)^{2j-2}
          \frac{\partial_{u_1} \partial_{u_2} Z(u_1=1,u_2=1)}
          {(\partial_{u_1}l_1) \, (\partial_{u_2}l_2)}
          \right|_{\chi=\frac{\pi}{2}} \nonumber \\
 P^{uu}_j &=&
   \left. \frac{1}{(2j)!}
          \left( \frac{\partial_\chi}{\partial_l \chi} \right)^{2j}
          Z(u_1=-1,u_2=-1) \right|_{\chi=\frac{\pi}{2}}
\end{eqnarray}
and since $P^{bu}_j = P^{ub}_j$ by symmetry, this suffices
determine all $P^{\alpha \beta}_j$. Note also that by an easy duality
argument we have $P^{bb}_{j+1} = P^{uu}_{j}$ for $j \ge 1$.

We find the following explicit results for $j \le 3$, here given
to order $\sim q^8$:
\begin{eqnarray}
  P_0 &=& 1-q^{\frac{1}{3}}-q^{\frac{4}{3}}+2q^2-2q^{\frac{7}{3}}+2q^3
   -2q^{\frac{10}{3}}+4q^4-4q^{\frac{13}{3}}+4q^5-5q^{\frac{16}{3}}+8q^6
  \nonumber \\
  & & -8q^{\frac{19}{3}}+8q^7-10q^{\frac{22}{3}}+14q^8+\cdots \nonumber \\
 P^{bb}_1 &=& q^{\frac{1}{3}}-2q+q^{\frac{4}{3}}-2q^2+2q^{\frac{7}{3}}
   -4q^3+6q^{\frac{10}{3}}-6q^4+8q^{\frac{13}{3}}-12q^5+13q^{\frac{16}{3}}
   -16q^6 \nonumber \\
   & & +20q^{\frac{19}{3}}-28q^7+30q^{\frac{22}{3}}-38q^8+\cdots \nonumber \\
 P^{ub}_1 &=& q-q^2-q^{\frac{10}{3}}-q^4-q^{\frac{13}{3}}+4q^5
   -2q^{\frac{16}{3}}+2q^6-3q^{\frac{19}{3}}+6q^7-5q^{\frac{22}{3}}
   +7q^8+\cdots \nonumber \\
 P^{uu}_1 &=& q^2+q^3-2q^{\frac{10}{3}}+2q^4-2q^{\frac{13}{3}}+q^5
   -4q^{\frac{16}{3}}+3q^6-6q^{\frac{19}{3}}+10q^7-10q^{\frac{22}{3}}
   +12q^8+\cdots \nonumber \\
 P^{ub}_2 &=& q^{\frac{10}{3}}+q^{\frac{13}{3}}-2q^5+2q^{\frac{16}{3}}
   -2q^6+3q^{\frac{19}{3}}-7q^7+5q^{\frac{22}{3}}-9q^8+\cdots \nonumber \\
 P^{uu}_2 &=& q^5+q^6+q^8+\cdots \nonumber \\
 P^{ub}_3 &=& q^7+q^8+\cdots
\end{eqnarray}
and $P^{uu}_3 = q^{\frac{28}{3}}+\cdots$. The evaluation of the
complete series for aspect ratio $\tau=1$ leads to the following
numerical values:

\medskip
\begin{tabular}{l|lllll}
$j$ & $\sum_{\alpha,\beta} P^{\alpha\beta}_j$ & $P^{\rm bb}_j$ &
       $P^{\rm ub}_j = P^{\rm bu}_j$ & $P^{\rm uu}_j$ \\ \hline
 0 & 0.6364540018880 & & \\
 1 & 0.3615910259567 & 0.2770671481561 & 0.0413139498152 & 0.0018959781702 \\
 2 & 0.0019548143402 & 0.0018959781702 & 0.0000293394720 & 0.0000001572261 \\
 3 & 0.0000001578149 & 0.0000001572261 & 0.0000000002943 & 0.0000000000002 \\
\end{tabular}
\medskip

\noindent
These values could presumably be verified by numerical
simulations in a square geometry.

\section{Conclusion}

In this article we have studied a densely packed loop model on the
annulus, with general loop weights that distinguish the two boundaries
and the homotopy class of the loops. The main result is the exact
seven-parameter continuum limit partition function (\ref{SuperGuess}).
We have verified that a range of special cases of this expression
agree with existing results in the literature, and used it to derive
new refined crossing probabilities in critical percolation.

The directions for future work are quite numerous \cite{DJS}. Let us
discuss briefly a few of them:

\begin{itemize}

\item Distinguishing both rims of the annulus by non-trivial boundary
  conditions is related with  properties of  $1BTL$ boundary
  condition changing operators. Resulting  fusion rules are encoded
  in the result (\ref{SuperGuess}). An intriguing---and to our
  knowledge novel---feature is that the fusion here depends on a
  parameter $n_{12}$ which is unrelated to those characterizing the
  two individual $1BTL$ operators.

\item Specializing the two-boundary model to simpler cases gives rise
  to a rich hierarchy of restrictions.  For instance, the two-boundary
  model with $n_{12} = n_1$ and $n_2 = n$ becomes the one-boundary
  model, and with $n_1=n$ this in turn becomes the standard
  (``zero-boundary'') Temperley-Lieb model. Moreover, in each case
  there are ``magical'' values of the weights, typically corresponding
  to one of the $r$-type parameters taking an integer value.
  Each of these restrictions corresponds to the disappearence of some
  of the states in the transfer matrix, the vanishing of certain
  eigenvalue amplitudes, and the reorganization of the Hilbert space
  into new modules. There is a rich algebraic meaning of this
  truncation hierarchy.

\item The present work pertains to the dense phase of the O($n$)
  model. In the dilute case the possibilities are richer: in addition
  to the boundary-specific $n$-type weights, one can weigh differently
  the boundary monomers depending on the type of loop to which they
  belong. This gives rise to several surface transitions. Some of
  those will be insensitive to the values of the $r$-type parameters,
  others will correspond to the usual swapping of indices (i.e.,
  $h_{r,s} \to h_{s,r}$ in the Kac formula), and yet others lead to
  genuinely new behaviour.

\end{itemize}

\subsection*{Acknowledgements}

We thank J.~Cardy, I.~Kostov, R.~Nepomechie and B.~Nienhuis for helpful comments
and for correspondence. This work was supported by the European
Community Network ENRAGE (grant MRTN-CT-2004-005616), by the Agence
Nationale de la Recherche (grant ANR-06-BLAN-0124-03), and by the ESF Network INSTANS.

%
%
%
%

\end{document}